\documentstyle[aps,psfig]{revtex}
\newcommand{\beq}{\begin{equation}}
\newcommand{\eeq}{\end{equation}}
\newcommand{\beqa}{\begin{eqnarray}}
\newcommand{\eeqa}{\end{eqnarray}}

\begin{document}
\preprint{LBNL-56660}
\title{Multiple Parton Scattering in Nuclei: Heavy Quark Energy Loss
and Modified Fragmentation Functions}
\author{Ben-Wei Zhang$^1$, Enke Wang$^{1,2}$ and Xin-Nian Wang$^2$}
\address{$^1$Institute of Particle Physics, Huazhong Normal University,
         Wuhan 430079, China}
\address{$^2$Nuclear Science Division, MS 70R0319,
Lawrence Berkeley National Laboratory, Berkeley, CA 94720 USA}

\maketitle

\vspace{-2.0in}
{\hfill LBNL-56660}
\vspace{2.in}

\begin{abstract}
\baselineskip=12pt Multiple scattering, induced radiative energy
loss and modified fragmentation functions of a heavy quark in
nuclear matter are studied within the framework of generalized
factorization in perturbative QCD. Modified heavy quark fragmentation
functions and energy loss are derived in detail with
illustration of the mass dependencies of the Landau-Pomeranchuk-Migdal
interference effects and heavy quark energy loss. Due to the quark mass
dependence of the gluon formation time, the nuclear size dependencies
of nuclear modification of the heavy quark fragmentation function
and heavy quark energy loss are found to change from a linear to a
quadratic form when the initial energy and momentum scale are
increased relative to the quark mass. The radiative energy loss of
the heavy quark is also significantly suppressed due to limited
cone of gluon radiation imposed by the mass. Medium
modification of the heavy quark fragmentation functions is
found to be limited to the large $z$ region due to the form
of heavy quark fragmentation functions in vacuum.

\end{abstract}


\baselineskip=16pt

\section{Introduction}

In ultra-relativistic heavy-ion collision, energetic partons from
initial hard processes have to propagate through the produced
dense medium and therefore suffer multiple scattering and lose a
significant amount of energy. Such energy loss or jet quenching
has been proposed as a good probe of the hot and dense medium
formed in high-energy nuclear collisions \cite{GP,WG}. It in
effect suppresses the final leading hadron distribution from the
propagating parton giving rise to modified fragmentation functions
and the final hadron spectra \cite{WH,cw}. Recent theoretical
studies \cite{GW1,BDMPS,Zak,GLV,Wie} all show that the effective
parton energy loss is proportional to the gluon density of the
medium. Therefore measurements of the parton energy loss will
enable one to extract the initial gluon density of the produced
hot medium. Strong suppression of high transverse momentum hadron
spectra is indeed observed by experiments
\cite{Phenix,Star,Phobos,Jacobs:2004qv} at the Relativistic
Heavy-Ion Collider (RHIC) at the Brookhaven National Laboratory
(BNL). The suppression pattern agrees very well with the jet
quenching mechanism \cite{gvwz,wang03}, indicating large parton
energy loss in a medium with large initial gluon density.
Comparing to jet quenching as measured in deeply inelastic
scattering off nuclei, the initial gluon density in central
$Au+Au$ collisions at $\sqrt{s}=200$ GeV is about 30 times higher
than that in a cold $Au$ nuclei \cite{EW1}. Such a high initial
gluon density is unprecedented and is a strong indication of the
formation of quark gluon plasma.

The extraction of the initial gluon density from jet quenching
pattern as measured at RHIC relies on the assumption that it is
caused by parton multiple scattering and induced radiation. While
such an assumption is based on a solid physical picture and is
supported by a multitude of experimental data \cite{Wang:2003aw},
it is still important to have additional and independent study
of the consequences of parton energy loss. Quenching of heavy
quark spectra has been proposed as a special probe
because of the unique mass dependence of the energy loss
and medium modification of the fragmentation
functions \cite{DK,DG:charm,zww,Armesto}. In this paper, we
will apply the framework of twist expansion that was
developed for the study of medium modification of parton
fragmentation functions to heavy quarks.

Multiple parton scattering inside nuclei in generally is a higher
twist process that involves multiple parton correlations. By
dimensional counting, such higher twist processes are
power-suppressed in terms of the momentum scale $Q^2$ involved.
Therefore, one can have a systematic expansion of the cross
section in $1/Q^2$. Since the probability of multiple scattering
increases with the nuclear size, the leading higher twist
contribution should be enhanced by $A^{1/3}$. Furthermore,
multiple parton correlations involve only the intrinsic properties
of the nuclei. They should be independent of the hard processes
involved. This twist expansion is referred to as generalized
factorization \cite{LQS}. Such framework has been applied to
semi-inclusive DIS on nuclear targets to study nuclear
modification of light (massless) quark fragmentation functions and
effective parton energy loss \cite{GW}. Because of the
Landau-Pomeranchuk-Migdal interference \cite{LPM}, the phase space
available for the induced gluon radiation is limited that is also
proportional to $A^{1/3}$. The final twist-four contribution to
the modified fragmentation functions is then proportional to
$A^{2/3}/Q^2$ \cite{GW}. Such a quadratic dependence on nuclear
size is indeed observed in semi-inclusive deep inelastic
lepton-nucleus experiments \cite{EW1,hermes}.

In this paper, we will extend the study of
modified fragmentation functions in nuclei \cite{GW,ZW}
to heavy quarks when they propagate through nuclear matter.
We will derive the modified heavy quark fragmentation functions
and the effective energy loss. To demonstrate
the effect of quark mass, we will compare the results
with the ones for light quarks. One of the most important
effect is the reduction of gluon formation time when it
is radiated from a slow heavy quark whose virtuality is
not much larger than its mass.
Such a reduction will effectively eliminate LPM effect
and the nuclear size dependence of the modification will become
linear in contrast to the case of a light quark.
The second mass effect is the significant reduction of
induced quark energy loss due to limited gluon radiation angle
imposed by the mass. With detailed data analysis of experimental
data both in the single electron channel \cite{phenix2} and
direct measurement of heavy mesons \cite{star-charm}, one could
learn more about the parton energy loss mechanism in dense
matter.

The results of heavy quark energy loss in the present
study were already reported in Ref.~\cite{zww}.
In this work, we will elaborate the detailed
derivation and focus on numerical calculations and
discussions about modified heavy quark fragmentation functions.
The paper is organized as follows. In
the next section we will present the theoretical formalism of our
calculation including the generalized factorization of twist-4
processes. In Section III we will show in detail the
calculation of different contributions to the modified heavy quark
fragmentation function and energy loss due to multiple scattering.
In Section IV we will numerically evaluate and discuss the modified
fragmentation functions of a heavy quark propagating in nuclei.
In Section V, discussion and numerical calculation of heavy quark
energy loss will be given. We will demonstrate the mass effects by
discussing how the dependence on medium size changes from a
linear to a quadratic dependence when the energy of the heavy
quark and the momentum scale is increased, and the suppression of
the energy loss for the heavy quark relative to a light quark
due to ``dead-cone'' effect \cite{DK}. Section VI will
summarize our work.


\section{Generalized Factorization}
In order to separate the complication of heavy quark production
and propagation, we consider a simple process of charm quark
production via the charge-current interaction in DIS off a large
nucleus. The results can be easily extended to heavy quark
propagation in a hot medium. The differential cross section for
the semi-inclusive process $\ell(L_1) + A(p) \longrightarrow
\nu_{\ell}(L_2) + H(\ell_H) +X$ can be expressed as
\begin{equation}
E_{L_2}E_{\ell_H}\frac{d\sigma_{\rm DIS}}{d^3L_2d^3\ell_H}
=\frac{G_{\rm F}^2}{(4\pi)^3 s} L_{\mu\nu}^{cc}
E_{\ell_H}\frac{dW^{\mu\nu}}{d^3\ell_H} \; . \label{sigma}
\end{equation}
Here $L_1$ and $L_2$ are the four momenta of the incoming lepton
and the outgoing neutrino, $\ell_H$ the observed heavy  meson
momentum, $p = [p^+,m_N^2/2p^+,{\bf 0}_\perp]
\label{eq:frame}$ is the momentum per nucleon in the nucleus,
$m_N$ is the mass of nucleon and $s=(p+L_1)^2$. $G_{\rm F}$ is
the four-fermion coupling constant, and $q =L_2-L_1 = [-Q^2/2q^-,
q^-, {\bf 0}_\perp]$ the momentum transfer via the exchange of a
vector boson $B(q)$. The charge-current leptonic tensor is given by
$L_{\mu\nu}^{cc}=1/2{\rm Tr}({\not\!L_1}\gamma_{\mu}
(1-\gamma_5){\not\!L_2}(1+\gamma_5)\gamma_{\nu})$. We assume
$Q^2\ll M_W^2$. The semi-inclusive hadronic tensor is defined as,
\begin{eqnarray}
E_{\ell_H}\frac{dW_{\mu\nu}}{d^3\ell_H}&=& \frac{1}{2}\sum_X
\langle A|J^+_\mu|X,H\rangle
\langle X,H| J_\nu^{+\dagger}|A\rangle
2\pi \delta^4(q+p-p_X-\ell_H) \label{hadronic}
\end{eqnarray}
where $\sum_X$ runs over all possible final states and
$J^+_\mu=\sum_f \bar{\psi}_f \gamma_\mu V \psi_f$  is the hadronic
charged current. Here we define $V=(1-\gamma_5)V_{ij}$ and $V_{ij}$
represents the CKM flavor mixing matrix \cite{AOT}. We want to
clarify that the symbol $Q$ in this paper stands for both the heavy
quark flavor and the momentum scale of the exchanged vector boson.

We consider the process of DIS in which  $W^{\pm }$ collides with a
light quark $q$ with momentum $k=xp$ producing a heavy quark $Q$
with mass $M $ and momentum $\ell_Q$. The heavy quark then
fragments into a heavy quark flavored hadron.
In this paper, we will take the charm quark as an example
and the cases for other heavy quarks will be straightforward. In
order to investigate the energy spectrum of charm fragmentation we
define the Lorentz-invariants $z=\ell_Q^-/q^-,
z_H=\ell_H^-/\ell_c^-$. The leading-twist and lowest order
perturbative QCD (pQCD) calculation of heavy quark production gives
\begin{equation}
\frac{dW_{\mu\nu}^{S(0)}}{dz_H} = \sum_q \int dx f_q^A(x)
H^{(0)}_{\mu\nu}(x,p,q,M) D_{Q \rightarrow H}(z_H)\ ,
\label{eq:w-s}
\end{equation}
where $f_q^A(x)$ is the quark distribution and $D_{Q\rightarrow
H}(z_H)$ is the non-perturbative heavy quark  fragmentation
function in vacuum \cite{P,MN,KS,Kretzer}. The hard partonic part is
\begin{eqnarray}
H^{(0)}_{\mu\nu}(k,q,M)&=& \frac{e_q^2}{2}\, {\rm Tr}( x\!\!\not\! p
\gamma_{\mu} V (\not\!q+x\!\!\not\!p ) V^{\dagger} \gamma_{\nu}) \,
\frac{2\pi}{2p\cdot q} \delta(x-x_B-x_M) \, ,
\label{H0} \\
x_M&=&\frac{M^2}{2p^+q^-}\, , \, \, \,\, \, x_B =\frac{Q^2}{2p^+
q^-} \,\, . \label{xm}
\end{eqnarray}

In the case when the momentum scale $Q$ is much larger than the
heavy quark mass, large logarithms such as $\log(Q^2/M^2)$ arise
to all orders of the perturbative expansion, so the fix-order
perturbation theory breaks down and a perturbative resummation of
large quasi-collinear logs, $\log(Q^2/M^2)$,  should be performed
\cite{MN,KS}, which will give the corresponding QCD evolution
equations for parton distribution functions and heavy quark
fragmentation functions. After considering higher order
contributions, the inclusive tensor can be written as
\cite{MN,CGRT,CG,BKK,KKS} \beq \frac{dW^S_{\mu\nu}}{dz_H} = \sum_q
\int dx f_q^A(x,\mu^2) H^{(0)}_{\mu\nu}(x,p,q,M) D_{Q\rightarrow
H}(z_H,\mu^2)\, \label{factorize-H1}, \eeq where $D_{Q\rightarrow
H}(z_H,\mu^2)$ satisfies the Dokshitzer-Gribov-Lipatov-Altarelli-Parisi
(DGLAP) \cite{AP} QCD evolution equations.

In a nuclear medium, the propagating heavy quark
in DIS will experience additional scattering with other partons
from the nucleus. The rescattering may induce additional gluon
radiation and cause the heavy quark to lose energy. Such induced
gluon radiation will effectively give rise to additional terms in
the evolution equation leading to the modification of the
heavy quark fragmentation functions in a medium. These are
higher-twist corrections since they involve higher-twist parton
matrix elements and are power-suppressed. We will consider those
contributions that involve two-parton correlations from two
different nucleons inside the nucleus. They are proportional to
the size of the nucleus \cite{OW} and thus are enhanced by a
nuclear factor $A^{1/3}$ as compared to two-parton correlations in
a nucleon. As in previous studies \cite{GW}, we will neglect those
contributions that are not enhanced by the nuclear medium.

\begin{figure}
\centerline{\psfig{file=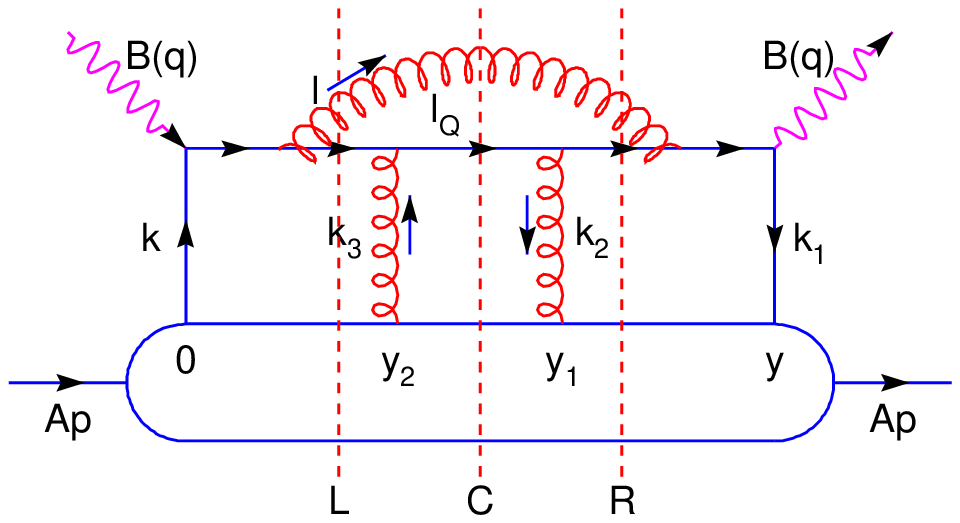,width=3in,height=2.0in}}
\caption{A typical diagram for quark-gluon re-scattering processes
with three possible cuts, central(C), left(L) and right(R).}
\label{fig1}
\end{figure}

For the production of heavy quarks there are usually two kinds of
mechanisms: intrinsic and extrinsic (via gluon fusion) heavy
quark production \cite{Nee}. Since we are only
interested in rescattering and induced gluon radiation of the
heavy quark after it is produced in DIS we will only consider a
simple case of intrinsic heavy quark production via
charged-current weak interaction. In this case, after small
modification we can still employ the generalized factorization of
multiple scattering processes \cite{LQS}. In this formalism with
collinear approximation, the double scattering contribution to
radiative correction from processes like the one illustrated in
Fig.~\ref{fig1} can be written in the following form,
\begin{eqnarray}
\frac{dW_{\mu\nu}^D}{dz_H} &=&
\sum_q \,\int_{z_H}^1\frac{dz}{z}D_{Q\rightarrow H}(z_H/z)
\int \frac{dy^-}{2\pi}\, dy_1^-\, dy_2^-\,\frac{d^2y_T}{(2\pi)^2}
d^2k_T \overline{H}^D_{\mu\nu}(y^-,y_1^-,y^-_2,k_T,p,q,M,z) \nonumber \\
&\ \times & e^{- i\vec{k}_T \cdot\vec{y}_T}\frac{1}{2} \,\langle A |
 \bar{\psi}_q(0)\, \gamma^+ \,A^+(y_{2}^{-},0_{T})\,
          A^+(y_{1}^{-},y_{T})\, \psi_q(y^{-}) |A\rangle \ .
\label{factorize-0}
\end{eqnarray}
Here
$\overline{H}^D_{\mu\nu}(y^-,y_1^-,y^-_2,k_T,p,q,M,z)$ is the
Fourier transform of the partonic hard part
$\widetilde{H}_{\mu\nu}(x,x_1,x_2,k_T,p,q,M,z)$ in momentum space,
\begin{eqnarray}
\overline{H}^D_{\mu\nu}(y^-,y_1^-,y^-_2,k_T,p,q,M,z) &=& \int dx\,
\frac{dx_1}{2\pi}\, \frac{dx_2}{2\pi}\,
       e^{ix_1p^+y^- + ix_2p^+y_1^- + i(x-x_1-x_2)p^+y_2^-}\nonumber \\
&\ & \times \widetilde{H}^D_{\mu\nu}(x,x_1,x_2,k_T,p,q,M,z) ,
\label{factorize-1}
\end{eqnarray}
where $k=[xp^+, 0, \vec{0}_\perp]$, $k_1=[x_1p^+, 0, \vec{0}_\perp]$,
$k_2=[x_2p^+, 0, \vec{k}_T]$, $k_3=[x_3p^+, 0, \vec{k}_T]$,
and $k_T$ is the relative transverse momentum carried by the
second parton in the double scattering.
We assume that $k_T$ is small and therefore can make collinear
expansion of the hard partonic cross section with respect to
the transverse momentum of the initial partons.
The first term in the collinear
expansion gives the eikonal contribution to the leading-twist
results, making the matrix element in the single scattering
process gauge invariant, while the second (or linear) term
vanishes for unpolarized initial and final states after
integration over $k_T$. The leading term in the collinear
expansion that contributes to the double scattering process
comes from the quadratic term in the collinear expansion,
\begin{eqnarray}
\frac{dW_{\mu\nu}^D}{dz_H} &=& \sum_q
\,\int_{z_H}^1\frac{dz}{z}D_{Q\rightarrow H}(z_H/z)
   \int \frac{dy^{-}}{2\pi}\, dy_1^-dy_2^-
\frac{1}{2}\,
     \langle A | \bar{\psi}_q(0)\,
           \gamma^+\, F_{\sigma}^{\ +}(y_{2}^{-})\,
 F^{+\sigma}(y_1^{-})\,\psi_q(y^{-})
     | A\rangle \nonumber \\
&\times &
    \left(-\frac{1}{2}g^{\alpha\beta}\right)
\left[\, \frac{1}{2}\, \frac{\partial^{2}}
                   {\partial k_{T}^{\alpha} \partial k_{T}^{\beta}}\,
    \overline{H}^D_{\mu\nu}(y^-,y_1^-,y^-_2,k_T,p,q,M,z)\, \right]_{k_T=0}\, .
\label{factorize}
\end{eqnarray}

There are many diagrams involving double parton scattering.
The hard part of the partonic scattering for each diagram,
$\widetilde{H}_{\mu\nu}(x,x_1,x_2,k_T,p,q,M,z)$, always contains
two $\delta$-functions from the on-shell conditions of the two
cut-propagators. These $\delta$-functions, together with the
contour integrations which contain different sets of poles in the
un-cut propagators, will determine the values of the momentum
fractions $x, x_1,$ and $x_2$ \cite{GW}. The phase factors in
$\overline{H}^D_{\mu\nu}(y^-,y_1^-,y^-_2,k_T,p,q,M,z)$
[Eq.~(\ref{factorize-1})] can then be factored out, which will be
combined with the partonic fields in Eq.~(\ref{factorize}) to form
twist-four partonic matrix elements or two-parton correlations.
The double scattering corrections in Eq.~(\ref{factorize}) can
then be factorized into the product of fragmentation functions,
twist-four partonic matrix elements and the partonic hard
scattering cross section.

\section{Double scattering and induced gluon radiation off a heavy quark}

According to the generalized factorization theorem in
Eq.~(\ref{factorize}), we should calculate the hard part of parton
multiple scattering. We will assume an axial gauge $n\cdot A=0$
with $n=[1,0^-,\vec{0}_\perp]$. The hard part of quark-gluon
double scattering has a total of 23 cut diagrams as illustrated in
Figs.~\ref{fig01}-\ref{fig11}.

We take the central cut diagram in Fig.~\ref{fig1} as an
example to show how to calculate the hard part. With the
conventional Feynman rule in the Standard Model one can write down
the hard partonic part of the central cut-diagram of
Fig.~\ref{fig1},
\begin{eqnarray}
\overline{H}^D_{C\,\mu\nu}(y^-,y_1^-,y_2^-,k_T,p,q,z)&=& \int
dx \frac{dx_1}{2\pi}\frac{dx_2}{2\pi} e^{ix_1p^+y^- +
ix_2p^+y_1^- + i(x-x_1-x_2)p^+y_2^-}
\int \frac{d^4\ell}{(2\pi)^4} \nonumber \\
&\times&\frac{{p^+}^2}{2}{\rm Tr}\left[x\!\!\not\!p \gamma_\mu V
n^\sigma n^\rho \widehat{H}_{\sigma\rho}V^{\dagger}\gamma_\nu
\right] 2\pi\delta_+(\ell^2)\, \delta(1-z-\frac{\ell^-}{q^-}) \; .
\label{eq:fig1-1}
\end{eqnarray}

\begin{eqnarray}
\widehat{H}_{\sigma\rho} &=&
\frac{C_F}{2N_c}g^4\frac{\gamma\cdot(q+k_1)+
M}{(q+k_1)^2-M^2-i\epsilon} \,\gamma_\alpha\,\frac{\gamma\cdot(q +
k_1-\ell)+M}{(q+k_1-\ell)^2-M^2-i\epsilon} \,\gamma_\sigma
(\gamma\cdot\ell_Q+M)\,\gamma_\rho
\nonumber \\
&\times &\varepsilon^{\alpha\beta}(\ell)\frac{\gamma\cdot(q+k
-\ell)+M} {(q+k-\ell)^2-M^2+i\epsilon} \,\gamma_\beta\,
\frac{\gamma\cdot(q + k)+M}{(q+k)^2-M^2+i\epsilon} \,\,2\pi
\delta_+(\ell_Q^2-M^2)\, , \label{eq:fig1-2}
\end{eqnarray}
where $k^2=k_1^2=0$, $p^2=m_N^2$, $\ell_Q^2=M^2$, $m_N$ and
$M$ are the nucleon mass and the produced heavy quark mass,
respectively. In addition, $\varepsilon^{\alpha\beta}(\ell)$ is
the polarization tensor of a gluon propagator in the axial gauge
and $\ell$, $\ell_Q=q+k_1+k_2-\ell$ are the 4-momenta carried by
the gluon and the final heavy quark, respectively. $z=\ell_Q^-/q^-$ is
the fraction of longitudinal momentum (the large minus component)
carried by the final heavy quark after gluon radiation.

In order to simplify the calculation of the trace part and extract
the leading contribution in the limit $\ell_T\rightarrow 0$ and
$k_T\rightarrow 0$, we also apply the collinear approximation to
complete the trace of the product of $\gamma$-matrices as
\begin{equation}
n^\sigma\widehat{H}_{\sigma\rho}n^\rho \approx
\frac{\gamma\cdot\ell_Q+M}{4\ell_Q^-} {\rm Tr} \left[\gamma^-
n^\sigma\widehat{H}_{\sigma\rho}n^\rho\right] \; . \label{coll}
\end{equation}
Therefore we have
\begin{eqnarray}
& &\frac{1}{2} {\rm Tr}\left[x\!\!\not\!p\gamma_\mu V n^{\sigma}
n^{\rho} \widehat{H}_{\sigma\rho} V^{\dagger}\gamma_\nu
\right]   \nonumber \\
 &\approx& \frac{1}{2} {\rm Tr}\left[x\!\!\not\!p
\gamma_\mu V (\gamma\cdot\ell_Q+ M) V^{\dagger}\gamma_\nu
\right]\frac{1}{4\ell_Q^-} {\rm Tr} \left[\gamma^-
n^{\sigma}\widehat{H}_{\sigma\rho}n^{\rho} \right] \;
.\label{coll-1}
\end{eqnarray}
After carrying out momentum integration
in $x$, $x_1$, $x_2$ and $\ell^{\pm}$ with the help of contour
integration and $\delta$-functions, the partonic hard part can be
factorized into the product of the matrix $H^{(0)}_{\mu\nu}(k,q)$
of vector boson and quark scattering, and the
quark-gluon rescattering part $\overline{H}^D$,
\begin{equation}
\overline{H}^D_{\mu\nu}(y^-,y_1^-,y_2^-,k_T,p,q,M,z) = \int dx
H^{(0)}_{\mu\nu}(k,q,M)\
\overline{H}^D(y^-,y_1^-,y_2^-,k_T,x,p,q,M,z)\, ,\label{eq:hc0}
\end{equation}
where $ H^{(0)}_{\mu\nu}(k,q,M)$ is defined in Eq.~(\ref{H0}).
Contributions from all the diagrams have this factorized from.
Therefore, we need only list the rescattering part
$\overline{H}^D$ for different diagrams in the following.

We also define the momentum fractions
\begin{eqnarray}
  x_L&=&\frac{\ell_T^2}{2p^+q^-z(1-z)} \,\, ,\,\,
  x_D=\frac{k_T^2-2\vec{k}_T\cdot \vec{\ell}_T}{2p^+q^-z} \, ,
\label{xld}
\end{eqnarray}
as in the light quark case.
Then, for the central-cut diagram in Fig.~\ref{fig1} we have
\cite{GW},
\begin{eqnarray}
\overline{H}^D_{1,C }(y^-,y_1^-,y_2^-,k_T,x,p,q,M,z)&=& \int
d\ell_T^2\,
\frac{(1+z^2)\ell_T^2+(1-z)^4 M^2}{(1-z)(\ell_T^2+(1-z)^2M^2)^2} \nonumber \\
&\times&\frac{\alpha_s}{2\pi}\, C_F \,
 \frac{2\pi\alpha_s}{N_c}
\overline{I}_{1,C}(y^-,y_1^-,y_2^-,\ell_T,k_T,x,p,q,M,z)
 \, , \label{eq:hc1}
\end{eqnarray}
\begin{eqnarray}
\overline{I}_{1,C }(y^-,y_1^-,y_2^-,\ell_T,k_T,x,p,q,M,z)
&=&e^{i(x+x_L)p^+y^- + ix_Dp^+(y_1^- - y_2^-)}
\theta(-y_2^-)\theta(y^- - y_1^-) \nonumber \\
& &\hspace{-1.0in}\times \left[1-e^{-i(x_L+(1-z)x_M/z
)p^+y_2^-}\right]
\left[1-e^{-i(x_L+(1-z)x_M/z)p^+(y^- - y_1^-)}\right] \; .
\label{eq:Ic1}
\end{eqnarray}
The above contribution resembles the cross section of dipole
scattering and contains essentially four terms. The first diagonal
term corresponds to the so-called hard-soft process where the
gluon radiation is induced by the hard scattering between the
vector boson $B$ and an initial quark with momentum $k$. The quark
is knocked off-shell by the $B$ boson and becomes on-shell again
after radiating a gluon. Afterwards the on-shell quark (or the
radiated gluon) will have a secondary scattering with another soft
gluon from the nucleus. The second diagonal term is due to the
so-called double hard process where the quark is on-shell after
the first hard scattering with the vector boson. The gluon
radiation is then induced by the scattering of the quark with
another gluon that carries finite momentum fraction
$x_L+(1-z)x_M/z+x_D$. The two off-diagonal terms are
interferences between hard-soft and double hard processes.
The cancellation between the two diagonal and off-diagonal
terms essentially gives rise to the LPM interference for the
induced gluon radiation. From Eq.~(\ref{eq:Ic1}), we get the
formation time for gluon radiation from a heavy quark,
\begin{equation}
\tau_f^Q\equiv \frac{1}{(x_L+(1-z)x_M/z)p^+} \, . \label{Htime}
\end{equation}
 In the limit of collinear radiation ($x_L\rightarrow 0$) or when the
formation time of the gluon radiation, $\tau_f^Q$, is much larger
than the nuclear size, the two processes (soft-hard and double hard)
have destructive
interference, leading to the LPM interference effect \cite{LPM}.
It is interesting to note that the formation time of gluon radiation
off a heavy quark $\tau_f^Q$ is shorter than that off a light
quark \cite{GW}
\begin{equation}
\tau_f^q\equiv \frac{1}{x_Lp^+} \,  .
\end{equation}
This is simply because the formation time for gluon radiation
is only relative to the propagation of the quark. The velocity
of low energy heavy quarks is much smaller than that of a light
quark. The corresponding gluon formation time is also smaller.
One should then expect the LPM interference
effect to be significantly reduced for intermediate energy heavy
quarks. It can be shown that this phenomenon will also modify the
dependence of the heavy quark energy loss on the nuclear size,
which will be discussed in detail in Section V.

In addition to the central-cut diagram, we also should take into
account the asymmetrical cut-diagrams in Fig.~\ref{fig1}, which
represent interference between single and triple scattering. We
note that the trace part is the same as in the central-cut
diagram. The difference is just the phase factor. Thus we have
from these cut diagrams,


\begin{eqnarray}
\overline{I}_{1,L}(y^-,y_1^-,y_2^-,\ell_T,k_T,x,p,q,M,z)\,
&=&-e^{i(x+x_L)p^+y^- + ix_Dp^+(y_1^- - y_2^-)}
\theta(y_1^- - y_2^-)\theta(y^- - y_1^-) \nonumber \\
&\times &(1-e^{-i(x_L+(1-z)x_M/z)p^+(y^- - y_1^-)}) \, ,\label{eq:I1L} \\
\overline{I}_{1,R}(y^-,y_1^-,y_2^-,\ell_T,k_T,x,p,q,M,z)
&=&-e^{i(x+x_L)p^+y^- + ix_Dp^+(y_1^- - y_2^-)}
\theta(-y_2^-)\theta(y_2^- - y_1^-) \nonumber \\
&\times &(1-e^{-i(x_L+(1-z)x_M/z)p^+y_2^-}) \, .\label{eq:I1R}
\end{eqnarray}

With the same procedure we can obtain the contributions of all
other cut diagrams, which are listed in Appendix ~\ref{appa}.


\section{Modified Heavy Quark Fragmentation Function}
To calculate the leading twist-four contribution to the semi-inclusive
cross section according to the generalized factorization formula
in Eq.~(\ref{factorize-0}), one has to expand the hard partonic cross
section in $k_T$. We can rearrange different contributions
according to the $\theta$-functions in $\overline{H}^D_{C,R,L}$
(the sum of all the contributions from central-cut, right-cut or
left-cut diagrams) and define
\begin{eqnarray}
\overline{H}^D&=&\int d\ell_T^2 \frac{\alpha_s}{2\pi}
e^{i(x+x_L)p^+y^- +ix_Dp^+(y_1^- - y_2^-)}
\frac{2\pi\alpha_s}{N_c}
\nonumber \\
&\times&\left[H^D_C\theta(-y_2^-)\theta(y^--y_1^-)
-H^D_R\theta(-y_2^-)\theta(y_2^--y_1^-)
-H^D_L\theta(y^--y_1^-)\theta(y_1^--y_2^-)\right] \, .
\label{eq:htheta}
\end{eqnarray}
>From Eqs.~(\ref{eq:hc1})-(\ref{eq:hc11-2}), we can check that
\begin{equation}
\overline{H}^D_C(k_T=0)=\overline{H}^D_R(k_T=0)=\overline{H}^D_L(k_T=0)C_F \, \frac{(1+z^2)\ell_T^2+(1-z)^4 M^2}{(1-z)[\ell_T^2+ (1-z)^2
M^2] ^2} \, .
\end{equation}

Since  one can reorganize the $\theta$-functions as
\begin{eqnarray}
& &\int dy_1^-dy_2^-\left[\theta(-y_2^-)\theta(y_2^--y_1^-)
+\theta(y^--y_1^-)\theta(y_1^--y_2^-)-\theta(-y_2^-)\theta(y^--y_1^-)
\right] \nonumber \\
&=&\int_0^{y^-}dy_1^-\int_0^{y_1^-}dy_2^- \; , \label{eq:theta}
\end{eqnarray}
one finds that $\overline{H}^D(k_T=0)$ in Eq.~(\ref{factorize-0})
just gives the eikonal contribution to the next-leading-order
correction of single  scattering which can be gauged away since
it does not correspond to any physical double scattering.

The leading contributions to the quark-gluon rescattering
result from the quadratic term in the $k_T$ expansion of
$\overline{H}^D$,
\begin{eqnarray}
\nabla^2_{k_T}H^D_{C(L,R)}|_{k_T=0}&=&4 C_A
\frac{1+z^2}{1-z}\frac{ \ell_T^4 }{[\ell_T^2+(1-z)^2 M^2]^4}\,
\widetilde{H^D}_{C(L,R)}
+{\cal O}(x_B/Q^2\ell_T^2)\, ,\label{hd}  \\
\widetilde{H_{C}^D}&=& c_1(z,\ell_T^2,M^2)
(1-e^{-i(x_L+(1-z)x_M/z)p^+y_2^-})
(1-e^{-i(x_L+(1-z)x_M/z)p^+(y^--y_1^-)})  \nonumber  \\
&+&c_2(z,\ell_T^2, M^2)\left[e^{-i(x_L+(1-z)x_M/z)p^+y_2^-}
(1-e^{-i(x_L+(1-z)x_M/z)p^+(y^--y_1^-)}) \right. \nonumber  \\
&+& \left. e^{-i(x_L+(1-z)x_M/z)p^+(y^--y_1^-)}
(1-e^{-i(x_L+(1-z)x_M/z)p^+y_2^-})\right]  \nonumber  \\
&+&c_3(z,\ell_T^2,M^2)
e^{-i(x_L+(1-z)x_M/z)p^+(y^--y_1^-)}e^{-i(x_L+(1-z)x_M/z)p^+y_2^-}
\label{hcd} \\
\widetilde{H_C^L}&=&c_4(z,\ell_T^2,M^2)(e^{-i(x_L+(1-z)x_M/z)p^+(y^--y_1^-)}
-e^{-i(x_L+(1-z)x_M/z)p^+(y^--y_2^-)}) \nonumber \\
&+&c_5(z,\ell_T^2,M^2)(1 -e^{-i(x_L+(1-z)x_M/z)p^+(y^--y_1^-)})
\label{hld}  \\
\widetilde{H_C^R}&=&c_4(z,\ell_T^2,M^2)(e^{-i(x_L+(1-z)x_M/z)p^+y_2^-}
-e^{-i(x_L+(1-z)x_M/z)p^+ y_1^-)}) \nonumber \\
&+&c_5(z,\ell_T^2,M^2)(1 -e^{-i(x_L+(1-z)x_M/z)p^+ y_2^-}),
\label{hrd}
\end{eqnarray}
where the coefficient $c_i(z,\ell_T^2,M^2), i=1,2,3,4,5$
are polynomial functions of $M^2/\ell_T^2$,
\begin{eqnarray}
c_1(z,\ell_T^2, M^2)&=& 1+\frac{(1-z)^2(z^2-6z+1) }{1+z^2}
\frac{M^2}{\ell_T^2} +
\frac{2z(1-z)^4}{1+z^2}\frac{M^4}{\ell_T^4}  \, , \label{c1}\\
 c_2(z,\ell_T^2, M^2)&=&\frac{(1-z)}{2}
 \left\{1-\left[\frac{(1-z)(2z^3-5z+8z-1)}{(1+z^2)}\right.
 +\frac{2C_F}{C_A}(1-z)^3\right]\frac{M^2}{\ell_T^2}
\nonumber \\
&-&\left[\frac{z(1-z)^4(3z-1)}{(1+z^2)}\right.
\left.\left.+\frac{2C_F}{C_A}
 \frac{(1-z)^7}{(1+z^2)}\right]\frac{M^4}{\ell_T^4} \right\} \, ,
   \label{c2} \\
c_3(z,\ell_T^2, M^2)&=&\frac{C_F (1-z)^2}{C_A}
\left[1-\frac{8z(1-z)^2}{1+z^2}\frac{M^2}{\ell_T^2} \right. \left.
-\frac{(1-z)^4(z^2-4z+1)}{1+z^2}\frac{M^4}{\ell_T^4}
\right] \, , \label{c3} \\
c_4(z,\ell_T^2, M^2)&=&(1-z)^2\frac{M^2}{\ell_T^2}
\left[1+ \frac{(1-z)^4}{1+z^2}
M^2/\ell_T^2 \right]\left[\frac{C_F}{C_A} (1-z)^2 + 2z-1\right]
\, ,   \label{c4}  \\
c_5(z,\ell_T^2, M^2)&= & (1-z)^2\frac{M^2}{\ell_T^2}
\left[1+ \frac{(1-z)^4}{1+z^2}\frac{M^2}{\ell_T^2}\right] \, . \label{c5}
\end{eqnarray}
In the limit of $M^2=0$, one can recover from Eqs.~(\ref{hd})-(\ref{hrd})
the results for light quark multiple scattering with complete
calculation beyond helicity amplitude approximation \cite{ZW}.

Compared with the results of gluon radiation from
the light quark multiple scattering \cite{GW,ZW} which have
an overall form $1/\ell_T^4$, the radiative gluon spectrum
from a heavy quark in Eq.~(\ref{hd}) is suppressed by a factor
\begin{equation}
f_{Q/q}=\left[\frac{\ell_T^2}{\ell_T^2+(1-z)^2
M^2}\right]^4=\left[1+\frac{\theta_0^2}{\theta^2}\right]^{-4},
\label{f}
\end{equation}
where $\theta_0=M/q^-$ and $\theta=\ell_T/l^-$ is the angle of the
radiated gluon relative to the heavy quark. One can see that the
mass of the heavy quark provides a lower bound for the radiation
angle of the collinear gluon which dominates the gluon spectrum in
the case of radiation off a light quark. This effectively
suppresses the gluon radiation at angle smaller than the ratio of
the quark mass $M$ to its energy $q^-$, and thus reduces radiative
energy loss of a heavy quark. Such a suppression of small angle
gluons is often referred to as the ``dead cone'' effect \cite{DK}.
In the result of our current calculation, additional mass effects
on the final gluon spectrum are in the mass dependence of
coefficient functions $c_i(z,\ell_T^2,M^2)$, $i=1,2,3,4,5$ and
most importantly the mass dependence of the gluon formation time
[Eq.~(\ref{Htime})] which will dictate the LPM interference
pattern in induced gluon bremsstrahlung off a heavy quark. These
additional mass effects will significantly influence the final
radiative gluon spectra and modified the so-called ``dead cone''
effect. In fact, the coefficient functions $c_i(z,\ell_T^2,M^2)$
contain terms like $\frac{M^4}{\ell_T^4}$. Multiplied with
the factor in Eq.~(\ref{f}), they give finite contribution
to the gluon spectra at $\ell_T=0$. This in effect will fill
up the small angle cone along the direction of the propagating
heavy quark with soft gluons, as also pointed out in Ref.\cite{Armesto}.
The net ``dead cone'' effect still results in the significant
reduction of the heavy quark energy loss, as numerical results
shown in Fig.~\ref{q-1} and Fig.~\ref{xb-1} of Section V below.

Substituting Eqs.~(\ref{hd})-(\ref{hld}) in Eq.~(\ref{factorize}),
we have the leading higher-twist contribution to the semi-inclusive
tensor of heavy quark fragmentation in DIS off a nucleus,
\begin{eqnarray}
\frac{W_{\mu\nu}^{D}}{dz_H} &=&\sum \,\int dx H^{(0)}_{\mu\nu}
\int_{z_H}^1\frac{dz}{z}D_{Q\rightarrow H}(\frac{z_H}{z})
\frac{C_A \alpha_s}{2\pi}  \frac{1+z^2}{1-z} \int
\frac{d\ell_T^2}{[\ell_T^2+(1-z)^2 M^2]^4}\ell_T^4
\nonumber \\
&\times& \frac{2\pi\alpha_s}{N_c} T^{A,Q}_{qg}(x,x_L,M^2)
+ (g-{\rm fragmentation})+({\rm virtual\,\, corrections})\, ,
\label{wd1}
\end{eqnarray}
where
\begin{eqnarray}
T^{A,Q}_{qg}(x,x_L,M^2)&\equiv&T^{A,C}_{qg}(x,x_L,M^2)+T^{A,L}_{qg}(x,x_L,M^2)
+T^{A,R}_{qg}(x,x_L,M^2) \,\, ,
\label{TH}  \\
 T^{A,C}_{qg}(x,x_L,M^2)&=& \int
\frac{dy^{-}}{2\pi}\, dy_1^-dy_2^- \widetilde{H_C^D}\,
\frac{1}{2}\langle A | \bar{\psi}_q(0)\, \gamma^+\, F_{\sigma}^{\
+}(y_{2}^{-})\, F^{+\sigma}(y_1^{-})\,\psi_q(y^{-})
| A\rangle \nonumber \\
&\times & e^{i(x+x_L)p^+y^-}
 \theta(-y_2^-)\theta(y^- -y_1^-) \;\; , \label{TC} \\
T^{A,L}_{qg}(x,x_L,M^2)&=& \int \frac{dy^{-}}{2\pi}\, dy_1^-dy_2^-
\widetilde{H_L^D}\, \frac{1}{2}\langle A | \bar{\psi}_q(0)\,
\gamma^+\, F_{\sigma}^{\ +}(y_{2}^{-})\,
F^{+\sigma}(y_1^{-})\,\psi_q(y^{-})
| A\rangle \nonumber \\
&\times &  e^{i(x+x_L)p^+y^-}
\theta(y^- - y_1^-)\theta(y_1^- -y_2^-) \;\; , \label{TL} \\
T^{A,R}_{qg}(x,x_L,M^2)&=& \int \frac{dy^{-}}{2\pi}\, dy_1^-dy_2^-
\widetilde{H_R^D}\, \frac{1}{2}\langle A | \bar{\psi}_q(0)\,
\gamma^+\, F_{\sigma}^{\ +}(y_{2}^{-})\,
F^{+\sigma}(y_1^{-})\,\psi_q(y^{-})
| A\rangle \nonumber \\
&\times &  e^{i(x+x_L)p^+y^-}
\theta(-y_2^-)\theta(y_2^- -y_1^-) \;\; ,  \label{TR}
\end{eqnarray}
are twist-four quark-gluon correction functions inside the nucleus
with $\widetilde{H_C^D}$, $\widetilde{H_L^D}$ and $\widetilde{H_R^D}$
given in Eqs.~(\ref{hcd})-(\ref{hrd}).
They are all independent with each other because they involve
different $\theta$ functions. The twist-four parton
matrices $T^{A,L}_{qg}$ and $T^{A,R}_{qg}$ are new, which purely
result from the mass effect of the heavy quark and involve
left and right cut diagrams. These
new parton matrices will vanish when we take $M^2=0$ [See the
definitions in Eqs.~(\ref{hcd}-\ref{hrd})].
Furthermore, they are proportional to coefficients
$c_4(z,\ell_T^2,M^2)$ and $c_5(z,\ell_T^2,M^2)$ which
in turn contain an additional factor of $(1-z)^2$. In the
case of soft gluon radiation $z \rightarrow 1$, these matrix
elements are suppressed compared to the matrix elements involved
in the central cut diagrams.

During the collinear expansion, we have kept $\ell_T$ finite and
took the limit $k_T\rightarrow 0$. As a consequence, the gluon
field in one of the twist-four parton matrix elements in
Eqs.~(\ref{TC})-(\ref{TR}) carries zero momentum in the soft-hard
process. As argued in Ref.~\cite{GW}, this is due to the
omission of higher order terms in the collinear expansion. As a
remedy to the problem, a subset of the higher-twist terms in the
collinear expansion can be resummed to restore the phase factors
such as $\exp(ix_Tp^+y^-)$, where $x_T\equiv \langle
k_T^2\rangle/2p^+q^-z$ is related to the intrinsic transverse
momentum of the initial partons. Therefore, soft gluon fields in
the parton matrix elements will carry a fractional momentum $x_T$.

Until now we have only considered quark-gluon double scattering in a
nucleus. To make a complete calculation we should also take into
account the processes of quark-quark scattering. However, it has
been shown \cite{GW} that the contributions of quark-quark scattering are
suppressed by a factor $1/Q^2$ as compared with quark-gluon double
scattering. In the heavy quark case, it further
involves intrinsic heavy quark
distribution inside the nucleus which should be very small as
compared to light quark and gluon distributions. Thus, we
can completely neglect the contributions of quark-quark scattering
for heavy quark propagation.

The virtual
corrections in Eq.~(\ref{wd1}) can be obtained via unitarity
requirement similarly as in Ref.~\cite{GW}. Including these
virtual corrections and the single scattering contribution, we can
rewrite the semi-inclusive tensor in terms of a modified
fragmentation function $\widetilde{D}_{Q\rightarrow H}(z_H,\mu^2)$,
\begin{equation}
\frac{dW_{\mu\nu}}{dz_H}=\sum_q \int dx
\widetilde{f}_q^A(x,\mu_I^2) H^{(0)}_{\mu\nu}(x,p,q,M)
\widetilde{D}_{Q \rightarrow H}(z_H,\mu^2) \label{eq:Wtot}
\end{equation}
where $\widetilde{f}_q^A(x,\mu^2)$ is the quark
distribution function which in principle should also include the
higher-twist contribution \cite{MQiu} of the initial state
scattering. The modified effective heavy quark fragmentation
function is defined as
\begin{eqnarray}
\widetilde{D}_{Q\rightarrow H}(z_H,\mu^2)&\equiv&
D_{Q \rightarrow H}(z_H,\mu^2) \nonumber \\
&+&\int_0^{\mu^2} \frac{d\ell_T^2}{\ell_T^2+(1-z)^2 M^2}
\frac{\alpha_s}{2\pi} \int_{z_H}^1 \frac{dz}{z}
\Delta\gamma_{q\rightarrow qg}(z,x,x_L,\ell_T^2,M^2)
D_{Q\rightarrow H}(z_H/z)  \nonumber \\
&+&  \int_0^{\mu^2} \frac{d\ell_T^2}{\ell_T^2+z^2 M^2}
\frac{\alpha_s}{2\pi} \int_{z_H}^1 \frac{dz}{z}
\Delta\gamma_{q\rightarrow gq}(z,x,x_L,\ell_T^2,M^2)
D_{g\rightarrow H}(z_H/z) \, , \label{eq:MDq}
\end{eqnarray}
where $D_{Q\rightarrow H}(z_H,\mu^2)$ and
$D_{g\rightarrow H}(z_H,\mu^2)$ are the
leading-twist fragmentation functions of a heavy quark
in vacuum. The modified splitting functions are given as
\begin{eqnarray}
\Delta\gamma_{q\rightarrow qg}(z,x,x_L,\ell_T^2,M^2)&=&
\left[\frac{1+z^2}{(1-z)_+}T^{A,Q}_{qg}(x,x_L,M^2) +
\delta(1-z)\Delta
T^{A,Q}_{qg}(x,\ell_T^2,M^2) \right] \nonumber \\
&\times& \frac{2\pi C_A \alpha_s \ell_T^4} {[\ell_T^2+(1-z)^2
M^2]^3 N_c\widetilde{f}_q^A(x,\mu_I^2)}\, ,
\label{eq:r1}\\
\Delta\gamma_{q\rightarrow gq}(z,x,x_L,\ell_T^2,M^2)
&=& \Delta\gamma_{q\rightarrow qg}(1-z,x,x_L,\ell_T^2,M^2) \label{eq:r2}, \\
\Delta T^{A,Q}_{qg}(x,\ell_T^2,M^2) &\equiv & \int_0^1
dz\frac{1}{1-z}\left[ 2 T^{A,Q}_{qg}(x,x_L,M^2)|_{z=1} -(1+z^2)
T^{A,Q}_{qg}(x,x_L,M^2)\right] \, . \label{eq:delta-T}
\end{eqnarray}
The above equations are similar to the case of double scattering
of a light quark \cite{GW,ZW} except that the
splitting functions for gluon radiation of heavy quark are quite
different from the one in the light quark case.

To numerically calculate the modified heavy quark fragmentation
function and study the energy loss of a heavy quark we have to
estimate the twist-four parton matrices $T^{A,Q}_{qg}(x,x_L,M^2)$,
which are in principle not calculable and can only be measured
independently in experiments similarly as parton distribution
functions. Nevertheless, with some hypotheses they can be
factorized. Here we will apply the approximation adopted in Refs.
\cite{LQS,GW,OW},
\begin{eqnarray}
&&\int \frac{dy^{-}}{2\pi}\, dy_1^-dy_2^-
e^{ix_1p^+y^-+ix_2p^+(y_1^--y_2^-)}
\langle A | \bar{\psi}_q(0)\frac{\gamma^+}{2}
F_{\sigma}^{\ +}(y_{2}^{-}) F^{+\sigma}(y_1^{-}) \psi_q(y^{-})
| A\rangle
 \theta(-y_2^-)\theta(y^--y_1^-) \nonumber \\
& & \hspace{1.5in} \approx \frac{C}{x_A} f_q^A(x_1)\, x_2f_g^N(x_2)\, ,
\label{eq:t4matrix} \\
&& \int \frac{dy^{-}}{2\pi}\, dy_1^-dy_2^-
 e^{ix_1p^+y^-+ix_2p^+(y_1^--y_2^-) \pm i(x_L+(1-z)x_M/z)p^+y_2^-}
\nonumber  \\
& & \hspace{1.5in}\times \langle A | \bar{\psi}_q(0) \frac{\gamma^+}{2}
F_{\sigma}^{\ +}(y_{2}^{-}) F^{+\sigma}(y_1^{-}) \psi_q(y^{-})
| A\rangle \theta(-y_2^-)\theta(y^--y_1^-) \nonumber \\
&& \hspace{1.5in} \approx \frac{C}{x_A} f_q^A(x_1)\,
x_2f_g^N(x_2)e^{-(x_L+(1-z)x_M/z)^2/x_A^2}\, , \label{eq:off-mx}
\end{eqnarray}
where $x_A=1/m_N R_A$, $f_q^A(x)$ is the quark distribution inside
a nucleus, $f_g^N(x)$ is the gluon distribution inside a
nucleon  and $C$ is assumed to be a constant, reflecting the strength
of two-parton correlation inside a nucleus.

In soft radiation approximation, $z\rightarrow 1$, the
parton matrix elements $T^{A,L(R)}_{qg}(x,x_L,M^2)$ from left and
right cut diagrams are suppressed and thus can be neglected in
our following numerical calculation.
According to Eqs.~(\ref{eq:t4matrix}) and (\ref{eq:off-mx}), we
have
\begin{eqnarray}
T^{A,C}_{qg}(x,x_L,M^2)&\approx&\frac{c_1(z,\ell_T^2,M^2)\, C}{x_A}
(1-e^{-(x_L+(1-z)x_M/z)^2/x_A^2}) \left[f_q^A(x+x_L)\, x_Tf_g^N(x_T)\right.
\nonumber \\ &+&
\left. f_q^A(x-(1-z)x_M/z)(x_L+x_T+(1-z)x_M/z)f_g^N(x_L+x_T+(1-z)x_M/z)\right]
\nonumber \\
&+& \frac{c_2(z,\ell_T^2,M^2)\,C}{2x_A} \left\{
e^{-(x_L+(1-z)x_M/z)^2/x_A^2}\left[f_q^A(x+x_L)\, x_Tf_g^N(x_T)\right.\right.
\nonumber \\
&+& \left.
f_q^A(x-(1-z)x_M/z)(x_L+x_T+(1-z)x_M/z)f_g^N(x_L+x_T+(1-z)x_M/z)\right]
\nonumber  \\ &- & \left.
 2f_q^A(x-(1-z)x_M/z)(x_L+x_T+(1-z)x_M/z)f_g^N(x_L+x_T+(1-z)x_M/z) \right\}
 \nonumber \\
 &+&\frac{c_3(z,\ell_T^2,M^2)\,C}
 {x_A}f_q^A(x-(1-z)x_M/z)
 \nonumber \\
 & & \times(x_L+x_T+(1-z)x_M/z)f_g^N(x_L+x_T+(1-z)x_M/z) \,.
 \label{T-f}
\end{eqnarray}
To further simplify the
calculation, we assume $x_M(1-z)/z, x_L\ll x_T \ll x$. The modified parton
matrix elements can be approximated by
\begin{eqnarray}
& &T^{A,C}_{qg}(x,x_L,M^2) \nonumber \\
&\approx& \frac{\widetilde{C}}{x_A} \
 f_q^A(x)\left[(1-e^{-(x_L+(1-z)x_M/z)^2/x_A^2})a_1(z,\ell_T^2,M^2)
+ a_2(z,\ell_T^2,M^2)\right], \label{modT2}
\end{eqnarray}
where $\widetilde{C}\equiv 2C x_Tf^N_g(x_T)$ is a coefficient
which should in principle depends on $Q^2$ and $x_T$. Here we
will simply take it as a constant. In order to simplify notations
we have defined
\begin{eqnarray}
a_1(z,\ell_T^2,M^2)&=& c_1(z,\ell_T^2,M^2)-c_2(z,\ell_T^2,M^2) \,\,  ,
\label{eq:a1} \\
a_2(z,\ell_T^2,M^2)&=& c_3(z,\ell_T^2,M^2)/2 \,\,  .\label{eq:a2}
\end{eqnarray}

Under these approximations, the only parameter in our
calculation is $\widetilde{C}$ which is also the only
parameter that enters into the modified fragmentation functions
for light quarks \cite{GW}. In the study of experimental data
on modified light quark fragmentation in DIS off nuclear targets,
a value of this parameter $\widetilde{C}\simeq 0.0060$ was
found \cite{EW1} to describe the data very well. Such a
value is also consistent with that extracted from the study
of transverse momentum broadening of Drell-Yan processes
in $p+A$ collisions \cite{guo}. In our following numerical
calculations of nuclear modification of heavy quark fragmentation
functions we will take the same value.

According to Eq.~(\ref{eq:MDq}), both heavy quark and gluon
fragmentation functions contribute to the modified heavy quark
fragmentation function.
In order to include the contribution of gluon fragmentation, we
have to consider the fragmentation function for heavy quarks in
the next-to-leading order pQCD calculation. Here we follow the
ansatz in Ref.~\cite{CG} and express the overall fragmentation
function of a parton $i$ into the hadron $H$ as \cite{CGRT,CG}:
\beq
   D_{i\rightarrow H}(z_H,\mu)    \int_{z_H}^1 \frac{dz}{z} D_i^Q(z,\mu)
   D_{Q\rightarrow H}(z_H/z),
\label{ansatz} \eeq
where $D_i^Q(z,\mu)$ is the perturbative
fragmentation function (PFF) for a massless parton to fragment
into a massive heavy quark $Q$ within pQCD
cascade. The perturbative fragmentation function (PFF) satisfies
the normal DGLAP QCD evolution equations \cite{AP} and
$D_{Q\rightarrow H}(z_H/z)$ is a non-perturbative
fragmentation function, describing the transition from the heavy
quark to the heavy meson, {\it e.g.}, the non-perturbative charm quark
fragmentation function into $D$ meson in Eq.~(\ref{eq:Dc}).

>From the next-to-leading order pQCD calculations \cite{MN}, we can
extract the initial conditions of PFF's for the heavy quark at a
scale $\mu_0$ of the order of the heavy quark mass $M$ as
\begin{eqnarray}
&&D_Q^Q(z,\mu_0) = \delta(1-z) + {{\alpha_s(\mu_0)
C_F}\over{2\pi}}\left[
{{1+z^2}\over{1-z}}\left(\log{{\mu_0^2}\over{M^2}} -2\log(1-z)
-1\right)\right]_+ \, ,\label{DQQ} \\
&&D_g^Q(z,\mu_0) = {{\alpha_s(\mu_0) C_A}\over{2\pi}}[z^2 +
(1-z)^2]
\log{{\mu_0^2}\over{M^2}} \, ,\label{DgQ} \\
&&D_{q,\bar q,\bar Q}^Q(z,\mu_0) = 0 \, . \label{DqQ}
\end{eqnarray}
With these initial conditions for PFF's and the DGLAP evolution
equations we can obtain the PFF functions evolved up to any scale
$\mu >\mu_0$. After convoluting PFF with the non-perturbative
fragmentation function $ D_{Q\rightarrow H}(z_H/z)$ in
Eq.~(\ref{ansatz}) we can get the fragmentation function
$D_{i\rightarrow H}(z_H,\mu)$ in vacuum,
which can be applied to calculate the heavy quark production cross
section within QCD factorization formula such as
Eq.~(\ref{factorize-H1}) \cite{CG,Nee}.

The non-perturbative fragmentation
function for the charm quark to fragment into $D$ meson in vacuum
can be parameterized in Peterson type functional form \cite{P} as
\begin{equation}
D_{c\rightarrow D}(z)=\frac{N}{z[1-z^{-1}-\varepsilon_c/(1-z)]}\, ,
 \label{eq:Dc}
\end{equation}
where $ N$  normalizes $D_c(z)$ to $\int dz D_c(z)=1$. The
parameter $\varepsilon_c$ is related to the heavy quark mass ($M_c$
for charm quark) by $\varepsilon_c=\Lambda^2/M_c^2$ and
$\Lambda$ stands for a hadronic scale.

\begin{figure}
\centerline{\psfig{file=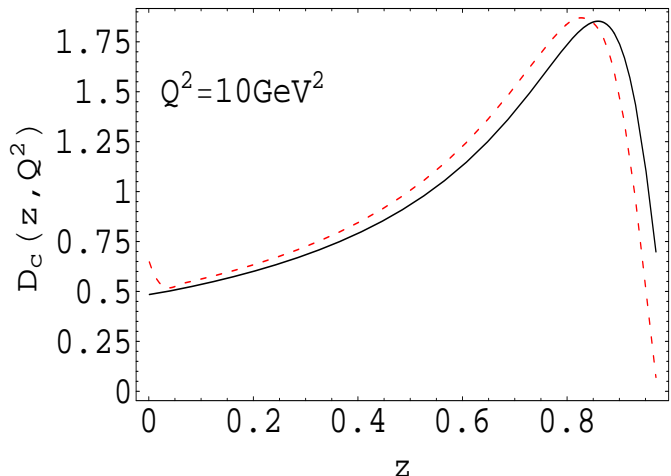,width=3.5in,height=2.5in}}
\caption{Charm quark fragmentation function into $D$ meson in
vacuum (solid line) and inside a nucleus (dashed line). For the
nuclear modification, $x_B=0.08$ and $x_A=0.05$ are used.}
\label{fig:cff}
\end{figure}

The PFF at scale $\mu > \mu_0 $ can be given by solving the DGLAP
equations, which is complicated in numerical calculations and
currently there is no parametrization forms available as for the
light quark fragmentation function \cite{BKK}. To simplify the
numerical calculations, we choose $\mu_0=Q$ as the first step for
numerical calculations and then obtain $D^D_c(z,Q)$ and
$D^D_g(z,Q)$ according to Eq.~(\ref{ansatz}). This approximation
is similar to the Approximation Case B used in Ref.~\cite{KKS}.
Shown in Fig.~\ref{fig:cff} as the solid line is the charm quark
fragmentation function into $D$ meson at $Q^2=10$ GeV$^2$ after
including higher order pQCD corrections. These vacuum heavy quark
fragmentation functions will be used as input in our numerical
computation of modified charm quark fragmentation function in
Eq.~(\ref{eq:MDq}). We note that in the limit of
vanishing quark masses the massless parton model expression
should be recovered in principle.

\begin{figure}
\centerline{\psfig{file=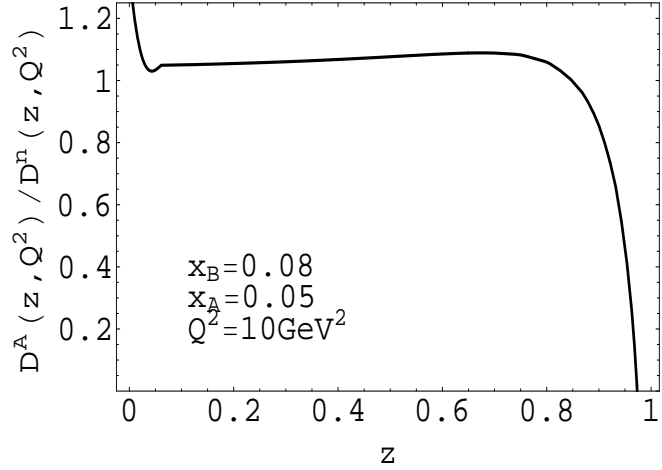,width=3.5in,height=2.5in}}
\caption{Modification factor for the charm quark fragmentation
function in a nucleus.} \label{mdf-fig}
\end{figure}

Shown in Fig.~\ref{fig:cff} as the dashed line is the modified
fragmentation function of a charm quark into the $D$ meson.
The value $x_A=0.05$ corresponds to a nucleus with a radius $R_A=4.25$ fm.
We have taken the charm quark mass $M=1.5$ GeV. One can see that
the modification due to the double scattering in a nucleus for heavy
quarks is quite different from light quarks \cite{EW1,GW}.
This is mainly caused by the form of heavy quark fragmentation
functions in vacuum which peak at large $z$. Because
of the multiple scattering and induced gluon radiation,
the position of the peak of the modified fragmentation function
is effectively shifted to a smaller value of $z$. As a
consequence, the heavy quark fragmentation function remains
unchanged, or even slightly enhanced for a large range of
fractional momentum $z$ as shown in Fig.~(\ref{mdf-fig}) by the
ratio of modified fragmentation function to the vacuum fragmentation
function. The modification only becomes significant and the
fragmentation function is suppressed at large $z$ above the
position of the peak. This is in sharp contrast
to the case of modified light quark fragmentation functions
which are suppressed relative to the vacuum form in a very
large range of $z$. Note that the heavy quark fragmentation
function is strongly enhanced at very small $z$, similarly to
the case of light quark fragmentation, due to heavy quark
pair production from the radiated gluons induced by multiple
scattering inside nuclei. However, this enhancement is
limited to much smaller $z$ than for light quarks fragmenting
into light hadrons.

\section{Heavy Quark Energy Loss}

Another advantage of studying multiple scattering of heavy quark
in medium is that one can actually measure the heavy quark
energy loss by flavor tagging, since the leading quark will
remain the heavy flavor which is very unlikely to be absorbed by
the medium. Similarly to the study of light quark energy loss,
we define  the heavy quark energy loss as the energy fraction
carried by the induced gluon,
\begin{eqnarray}
\langle\Delta z_g^Q \rangle(x_B,\mu^2) &=& \int_0^{\mu^2}d\ell_T^2
\int_0^1 dz \frac{\alpha_s}{2\pi}
 (1-z)\,\frac{\Delta\gamma_{q\rightarrow qg}(z,x_B,x_L,\ell_T^2)}
{\ell_T^2 +(1-z)^2 M^2} \nonumber \\
&=&\frac{C_A\alpha_s^2}{N_c} \int_0^{\mu^2}d\ell_T^2 \int_0^1 dz
\frac{(1+z^2)\ell_T^4}{[\ell_T^2 +(1-z)^2 M^2]^4}
\frac{T^{A,C}_{qg}(x_B,x_L,M^2)}{\widetilde{f}_q^A(x_B,\mu_I^2)}\,
. \label{eq:loss1}
\end{eqnarray}
Substituting the approximate expression for the nuclear twist-four
parton matrix in Eq.~(\ref{modT2}), we obtain,
\begin{eqnarray}
\langle\Delta z_g^Q\rangle(x_B,\mu^2) &=&\frac{ \widetilde{C}C_A
\alpha_s^2} {N_c \, x_A} \int_0^{Q^2} d\ell_T^2
\int_0^1 dz \frac{(1+z^2)\ell_T^4}  {[\ell_T^2 +(1-z)^2 M^2]^4 } \nonumber \\
 &\times & \left[(1-e^{-\widetilde{x}_L^2/x_A^2})a_1(z,\ell_T^2,M^2)
+a_2(z,\ell_T^2,M^2)\right] \, .
  \label{eq:H-loss-1}
\end{eqnarray}
Here we choose the factorization scale as $\mu^2=Q^2$ and define
$\widetilde{x}_L\equiv x_L+(1-z)x_M/z$. Note that the virtual
correction in $\Delta\gamma_{q\rightarrow qq}$ does not contribute
to the energy loss. Also, $\widetilde{x}_L/x_A=L_A^-/\tau_f$ with
$L_A^-=R_Am_N/p^+$ the nuclear size in the chosen frame. The
second term proportional to $a_2$ corresponds to a finite
contribution in the factorization limit. This term will survive in
the limit of complete LPM cancellation when double scattering acts
like a single scattering for induced gluon radiation. We have
neglected such a term in the study of light quark propagation
since it is proportional to $R_A$, as compared to the $R_A^2$
dependence from the first term due to the LPM effect. In this
study we have to keep the second term for heavy quark propagation
since the first term with the LPM interference effect will have a
similar nuclear dependence when the heavy quark mass reduces the
gluon formation time for low energy heavy quarks. However, for
energetic heavy quark, the mass can become negligible and one
should reach the limit of a light quark energy loss.

To elucidate the two different limits, we examine the phase factor
in Eq.~(\ref{eq:H-loss-1}),
\begin{eqnarray}
\frac{(x_L+(1-z)x_M/z)^2}{x_A^2} =\frac{x_B^2[\ell_T^2+(1-z)^2
M^2]^2}{ x_A^2 z^2(1-z)^2 Q^4}   \sim \frac{x_B^2 M^4}{ x_A^2 Q^4}
\, ,
\end{eqnarray}
and define it as $ T \equiv x_B^2 M^4/x_A^2 Q^4$, which
should control the LPM
interference effect, and therefore the behavior of the total heavy
quark energy loss. There are two distinct limiting behaviors of the
energy loss for different values of of $x_B$, $Q^2$ and $x_A$.

When $ T  \gg 1$ for $Q^2 \ll M^2$ or $x_B \gg x_A$, we have
$$1-e^{-\widetilde{x}_L^2/x_A^2} \simeq 1 \, ,$$
which means there is no LPM interference \cite{zww} and we obtain
\begin{eqnarray}
\langle\Delta z_g^Q\rangle \sim C_A \frac{ \widetilde{C}
\alpha_s^2}{N_c} \frac{x_B}{x_A Q^2} .
\end{eqnarray}
Since $x_A=1/m_N R_A$, the heavy quark energy loss in this case
depends linearly on the nuclear size $R_A$ as the Bethe-Heitler form
in Abelian gauge interaction. Similar results were also
derived in the generalized opacity expansion method \cite{DG:charm}.

In the opposite limit when $Q^2\gg M^2 $ or $x_B\ll x_A$, the
quark mass becomes negligible and $ T $ will take a moderate
value. The gluon formation time can  be much larger than the
nuclear size and therefore the LPM interference effect will
dominate again. In this case, one can make a variable change
$x_L\rightarrow \widetilde{x}_L$ with $\widetilde{x}_M=(1-z)x_M/z$
and $\widetilde{x}_{\mu}=\mu^2/2p^+q^-z(1-z)+\widetilde{x}_M$ in
Eq.~(\ref{eq:H-loss-1}) and obtain
\begin{eqnarray}
\langle\Delta z_g^H\rangle(x_B,\mu^2) &=&\frac{
\widetilde{C}C_A\alpha_s^2 x_B} {N_c Q^2 \, x_A}\int_0^1 dz
\frac{1+z^2}{z(1-z)} \int_{\widetilde{x}_M}^{\widetilde{x}_{\mu}}
d\widetilde{x}_L\frac{(\widetilde{x}_L-\widetilde{x}_M)^2}{
{\widetilde{x}_L}^4} \nonumber \\
&\times &
\left[(1-e^{-\widetilde{x}_L^2/x_A^2})a_1(z,\ell_T^2,M^2)
+a_2(z,\ell_T^2,M^2)\right]\, .    \label{eq:H-loss-2}
\end{eqnarray}
This form is very similar to the one for the light quark energy
loss \cite{ZW}. Since $a_1(z,\ell_T^2,M^2)$ and
$a_2(z,\ell_T^2,M^2)$ are dimensionless coefficients,  and  $ T$
have a moderate value, the exponential factor in the above equation
coming from the LPM interference regularizes the integration over
$\widetilde{x}_L$ and limits $\widetilde{x}_L <x_A$. We obtain
$\int d\widetilde{x}_L/ {\widetilde{x}_L}^2 \sim 1/x_A$. With
similar argument in Ref.~\cite{GW} we can conclude that the heavy quark
energy loss is proportional to
 \begin{equation}
\langle\Delta z_g^Q\rangle \sim C_A \frac{ \widetilde{C}
\alpha_s^2}{N_c} \frac{x_B}{x_A^2 Q^2} \; .
\end{equation}
Therefore, the heavy quark energy loss has a quadratic
dependence on the nuclear size when $Q^2\gg M^2$ or $x_B\ll x_A$.
This is understandable because under such condition, the
heavy quark becomes again relativistic and should behavior
like a light quark in terms of induced energy loss.

\begin{figure}
\centerline{\psfig{file=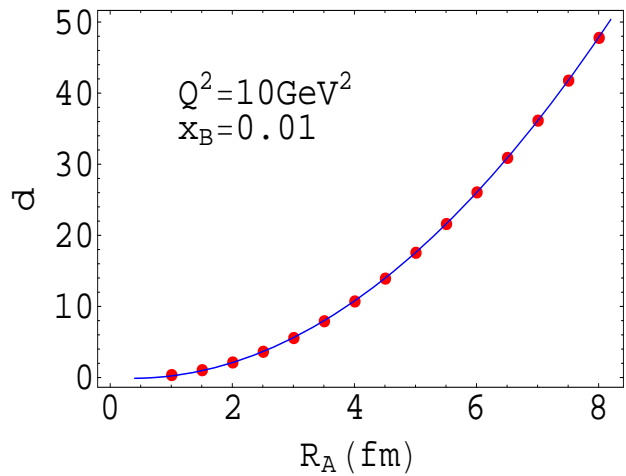,width=3.5in,height=2.5in}}
\caption{ The $R_A$  dependence of heavy quark energy loss for
a charm quark.} \label{xa-1}
\end{figure}
\begin{figure}
\centerline{\psfig{file=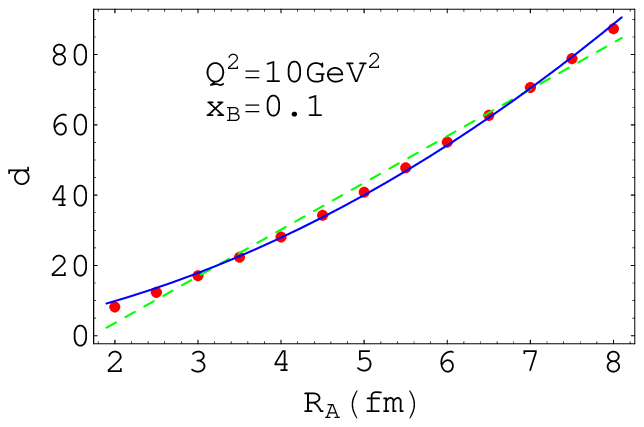,width=3.5in,height=2.5in}}
\caption{ The $R_A$ dependence of heavy quark energy loss for
a charm quark. } \label{xa-2}
\end{figure}
\begin{figure}
\centerline{\psfig{file=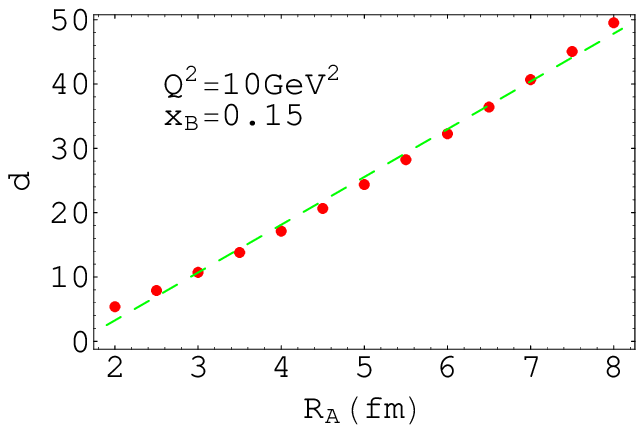,width=3.5in,height=2.5in}}
\caption{ The $R_A$  dependence of heavy quark energy loss for
a charm quark. } \label{xa-3}
\end{figure}
\begin{figure}
\centerline{\psfig{file=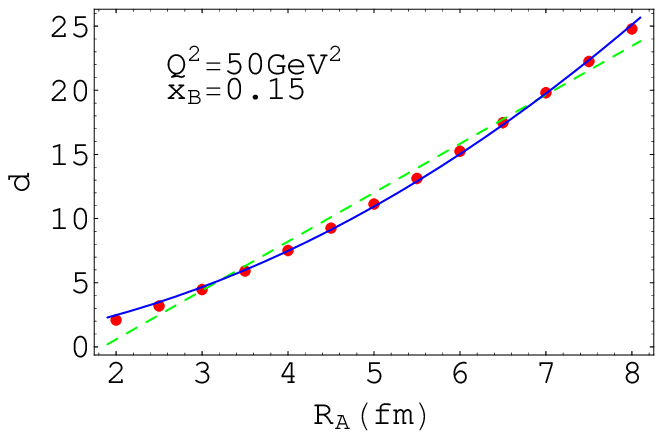,width=3.5in,height=2.5in}}
\caption{ The $R_A$ dependence of heavy quark energy loss for
a charm quark. } \label{xa-4}
\end{figure}
\begin{figure}
\centerline{\psfig{file=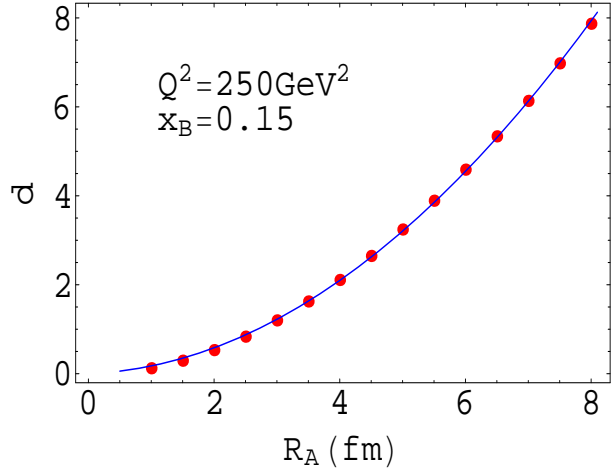,width=3.5in,height=2.5in}}
\caption{ The $R_A$ dependence of heavy quark energy loss for
a charm quark. } \label{xa-5}
\end{figure}

To present our numerical calculations of heavy quark energy loss,
we rescale the heavy quark energy loss by
$\widetilde{C}(Q^2)C_A\alpha_s^2(Q^2)/N_C$ and define
\begin{equation}
d=\langle\Delta
z_g^H\rangle\frac{N_C}{\widetilde{C}(Q^2)C_A\alpha_s^2(Q^2) }\,\,
.
\end{equation}
The $R_A$ dependence of the rescaled heavy quark energy loss is
shown in Figs.~\ref{xa-1}-\ref{xa-5}, where the points are
numerical results, the dashed lines are linear fit to the
numerical results and solid curves are quadratic fit. They clearly
show that the heavy (charm) quark energy loss has different
nuclear size dependencies as $Q^2$ and $x_B$ change. In
Fig.~\ref{xa-1}, Fig.~\ref{xa-2} and Fig.~\ref{xa-3} we fix
$Q^2=10$ GeV$^2$ and change the value of $x_B$ to demonstrate the
nuclear size $R_A$ dependence of heavy quark energy loss for a
charm quark ($M=1.5$ GeV). One notes that when $x_B$ is very small
(large heavy quark energy), the charm quark energy loss depends
quadratically on $R_A$ (see Fig.~\ref{xa-1}). However, as we
increase $x_B$ (decrease heavy quark energy) a gradual transition
from a quadratic dependence on $R_A$ of the energy loss to a
linear dependence (see Fig.~\ref{xa-2}) takes place. When $x_B$ is
very large the charm quark energy loss has an all most linear
dependence on nuclear size $R_A$ as Fig.~\ref{xa-3} illustrates.
Similarly, shown in Figs.~\ref{xa-3}, \ref{xa-4} and \ref{xa-5},
we see that even for large values of $x_B$ (where we fix
$x_B=0.15$) or small heavy quark energy, the same transition from
linear nuclear size dependence to quadratic dependence of the
heavy quark energy loss takes place as we increase $Q^2$.

Another mass effect on the induced gluon radiation is the
``dead-cone'' phenomenon \cite{DK} that suppresses the small
angle gluon radiation. Since the size of the dead-cone
$\theta_0=M/q^-$ [Eq.~(\ref{f})], within which the gluon radiation
is suppressed, is inversely proportional to the quark's energy,
the reduction of energy loss is stronger for a slower heavy quark.
For a heavy quark with either a high energy $q^-$ or virtuality
$Q^2$, its radiative energy loss should approach that of a light
quark. Since the dimensionless coefficients $a_1(z,\ell_T^2,M^2)$
and $a_2(z,\ell_T^2,M^2)$ in Eq.~(\ref{eq:H-loss-2}) also depend
on the heavy quark mass, they will have addition mass effects on
the heavy quark energy loss.

To illustrate the difference of energy loss between heavy quark
and light quark and the quark mass effect we define a ratio $R$ as:
\begin{equation}
R\equiv \frac{  \langle\Delta z_g^Q\rangle(x_B,\mu^2) } {
\langle\Delta z_g^q\rangle(x_B,\mu^2) } \,\, ,  \label{r}
\end{equation}
where $\langle\Delta z_g^q\rangle(x_B,\mu^2)$ is the light quark
energy loss \cite{ZW} which can be obtained by setting $M=0$ in
Eq.~(\ref{eq:H-loss-1}) and Eq.~(\ref{eq:H-loss-2}),
\begin{eqnarray}
\langle\Delta z_g^q\rangle(x_B,\mu^2) =\frac{
\widetilde{C}\alpha_s^2} {N_c x_A} \int_0^{Q^2}
\frac{d\ell_T^2}{\ell_T^4} \int_0^1 dz\, (1+z^2)
  (1-e^{-x_L^2/x_A^2})C_A[1-\frac{1-z}{2}]\, .
\label{eq:L-loss}
\end{eqnarray}

\begin{figure}
\centerline{\psfig{file=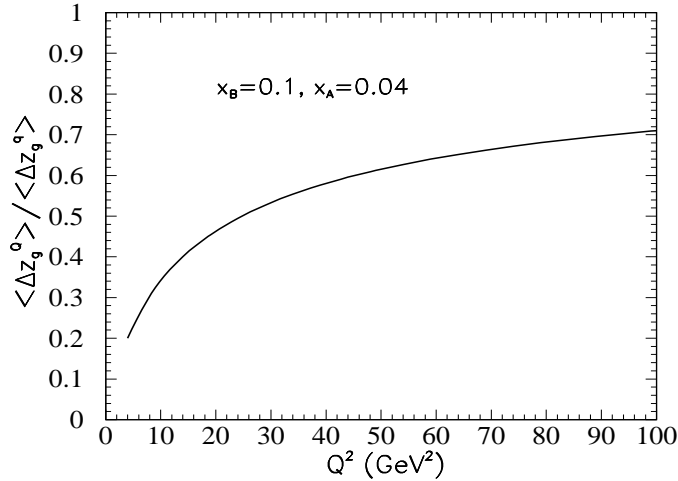,width=3.5in,height=2.5in}}
\caption{ The $Q^2$ dependence of R for a charm quark.} \label{q-1}
\end{figure}

In Fig.~\ref{q-1}, we show the change of the ratio $R$ with $Q^2$
for a charm quark ($M=1.5$ GeV) propagating in a nucleus with
$x_A=0.04$ and $x_B=0.1$. Please note that in the nuclear parton matrix
elements, the fractional momentum in a nucleon is limited to
$x_L<1$ due to the momentum conservation. Even though the Fermi
motion effect in a nucleus can allow $x_L>1$, the parton
distribution in this region is still significant suppressed. Thus
it provides a natural cut-off for $x_L$ in the numerical
integration over $z$ and $\ell_T$ in Eqs.~(\ref{eq:H-loss-1}) and
(\ref{eq:L-loss}).

We can observe that when $Q^2$ is not too large, the heavy quark
mass effect significantly suppresses the energy loss caused by
induced gluon radiation. When $M^2/Q^2 \rightarrow 0 $, the
effect of quark mass becomes negligible and $R\rightarrow 1$.
This is consistent with the pQCD factorization
theorem that when the momentum transfer is very large one can
neglect the effect of quark mass.

\begin{figure}
\centerline{\psfig{file=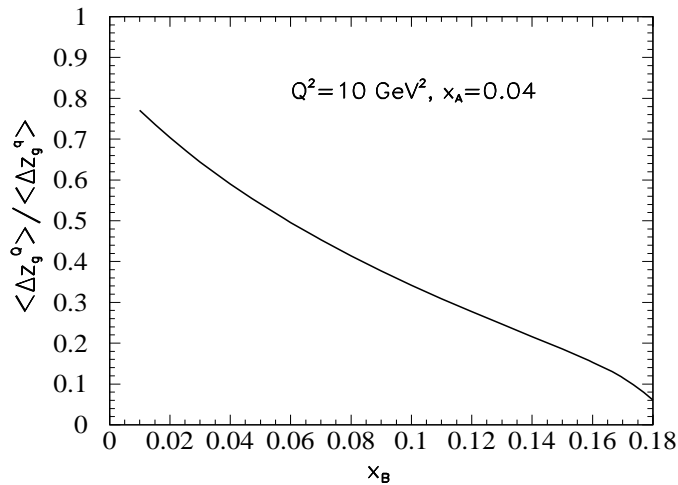,width=3.5in,height=2.5in}}
\caption{ The $x_B$ dependence of R for a charm quark. }
\label{xb-1}
\end{figure}

Shown in Fig.~\ref{xb-1} is the $x_B$ (or heavy quark energy)
dependence of the ratio between heavy quark and
light quark energy loss for fixed $Q^2=10$ GeV$^2$.
It is obvious that suppression of heavy quark energy loss
due to the ``dead-cone'' effect of heavy quark mass is
most significant when $x_B$ is large (or quark energy is small).
When $x_B$ is very small (quark energy is large), the
effect of quark mass is small and the quark energy loss
approaches that of a light quark.

\section{Summary}

Utilizing the generalized factorization of twist-four processes we
have studied the nuclear modification of heavy quark fragmentation
functions and the energy loss of a heavy quark propagating through
dense matter after it is produced via a hard process in DIS in the
twist expansion approach. Taking into account of the multiple
scattering suffered by the heavy quark we have derived the
modified heavy quark fragmentation functions related
to twist-four corrections with nuclear enhancement. We find that
the formation time for gluons radiated from a heavy quark is
smaller relative to that of a light quark, since it is
always measured against the propagation time of the quark.
With certain kinematics when the quark energy and virtuality
is small, the gluon formation time can become much smaller
than the nuclear size. In this case, the heavy quark energy
loss or the nuclear modification of the heavy quark
fragmentation functions have a linear nuclear size dependence.
We have shown through both analytic deduction and numerical
calculation that a gradual transition from a linear to a quadratic
nuclear size dependence takes place when one increases
the quark's energy or initial virtuality.

We also compared the energy loss of a heavy quark with that of a
light quark and demonstrate that the quark mass effect, including
the so-called ``dead-cone'' phenomenon, will  significantly
suppress the heavy quark energy loss when the momentum
transfer is not too large. This heavy quark mass effect will
decrease if the heavy quark energy, or the momentum scale $Q^2$ is
much larger than the quark mass. When $M\rightarrow 0$, our
calculations recover the results for massless quarks in previous
studies.

Similar to the case of light quark propagation \cite{EW1}, the
results discussed in this work can be easily extended to a hot and
dense medium, which will have practical consequences for heavy
quark production and suppression in heavy
ion collisions. When the data on direct measurement of $D$-meson
spectra in high-energy $A+A$ collisions become available in the near
future, one should be able to use the modified fragmentation
function in a parton model to study the modification of the
$D$-meson spectra \cite{Djordjevic:2004nq} and probe medium
properties similarly as one
has done for high $p_T$ light hadrons \cite{wang03}. The
different pattern of energy loss for heavy quarks, such as energy
and medium size dependence, will not only confirm the unique
feature of non-Abelian energy loss but also give more confidence
in using jet tomography to study properties of dense matter in
heavy-ion collisions.

\section*{Acknowledgements}
E. Wang thanks the hospitality of the Nuclear Theory Group
at Lawrence Berkeley National Laboratory during the completion
of this work.
This work was supported by NSFC under project Nos. 10347130,
10405011 and 10440420018, and by the Director, Office of Energy
Research, Office of High Energy and Nuclear Physics, Divisions of
Nuclear Physics, of the U.S. Department of Energy under Contract
No. DE-AC03-76SF00098.

\appendix

\section{}
\label{appa}

\begin{figure}
\centerline{\psfig{file=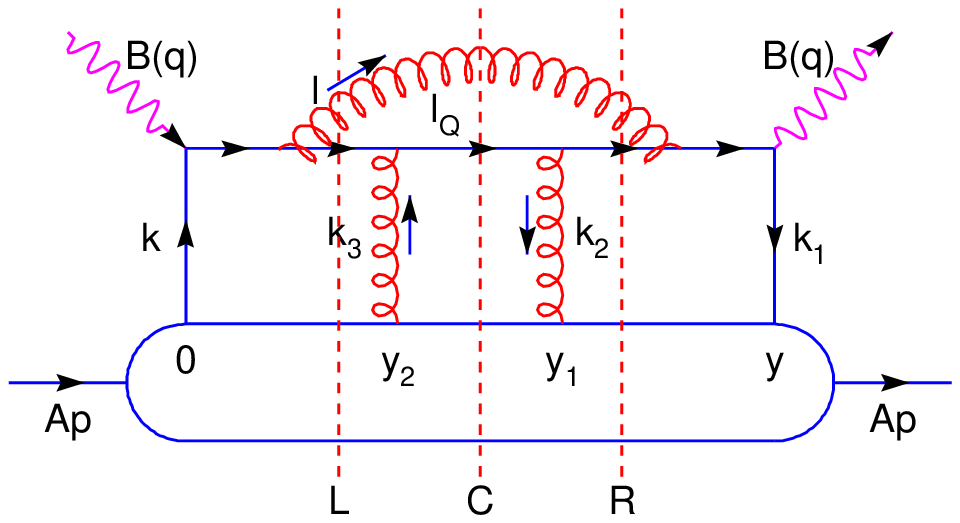,width=2.6in,height=1.8in}}
\caption{} \label{fig01}
\end{figure}

\begin{figure}
\centerline{\psfig{file=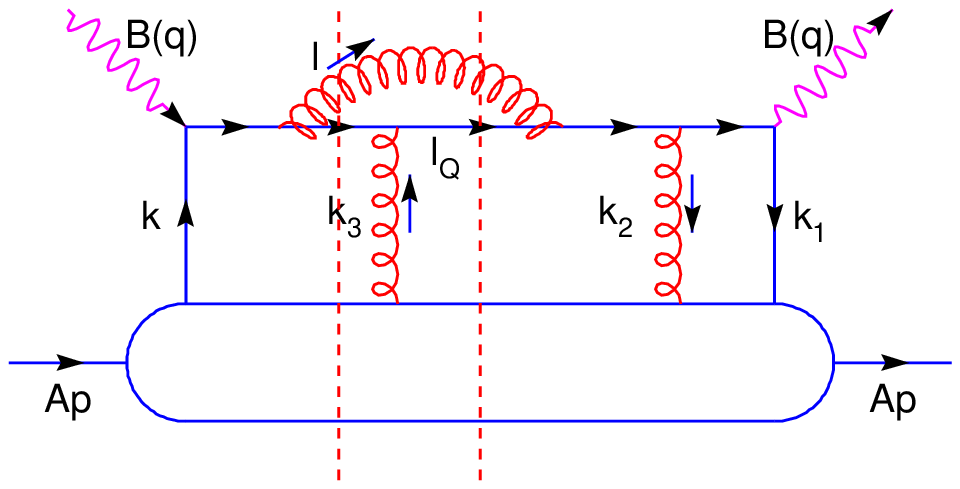,width=2.6in,height=1.8in}}
\caption{ } \label{fig2}
\end{figure}

\begin{figure}
\centerline{\psfig{file=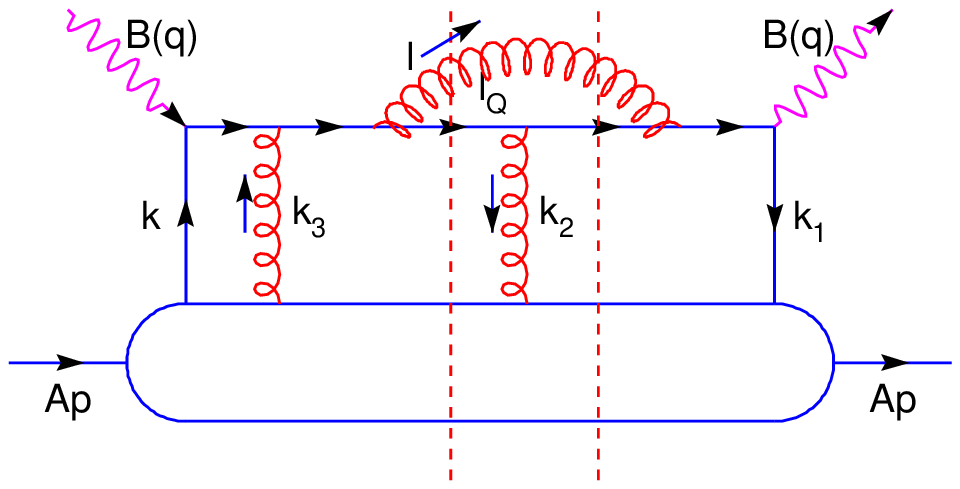,width=2.6in,height=1.8in}}
\caption{ } \label{fig3}
\end{figure}

\begin{figure}
\centerline{\psfig{file=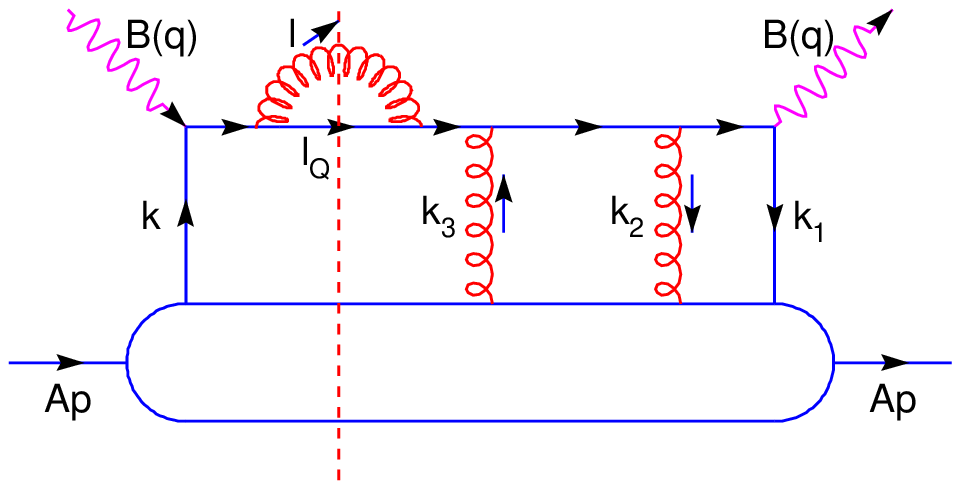,width=2.6in,height=1.8in}}
\caption{} \label{fig4}
\end{figure}

\begin{figure}
\centerline{\psfig{file=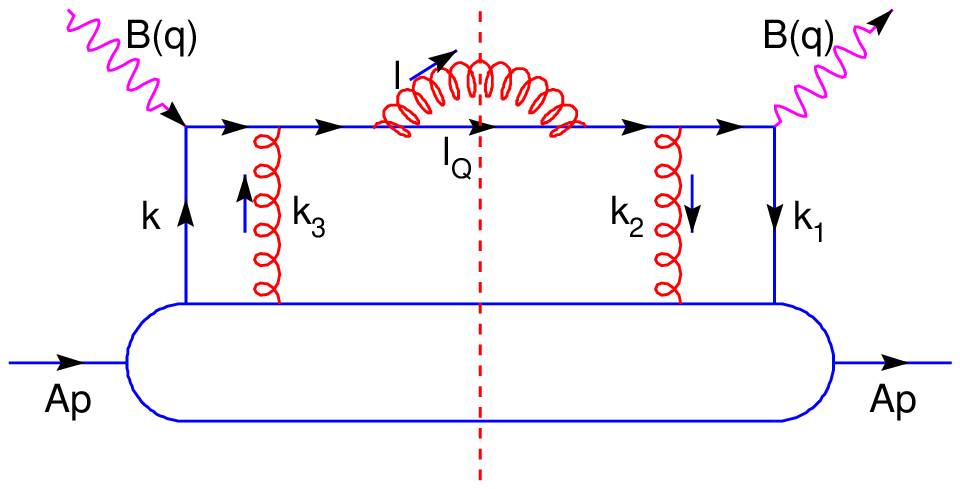,width=2.6in,height=1.8in}}
\caption{ } \label{fig5}
\end{figure}

\begin{figure}
\centerline{\psfig{file=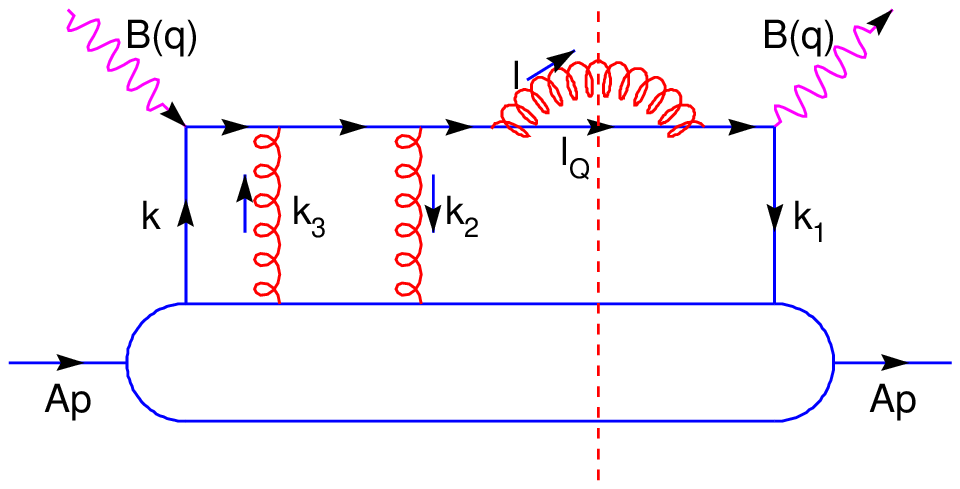,width=2.6in,height=1.8in}}
\caption{ } \label{fig6}
\end{figure}

\begin{figure}
\centerline{\psfig{file=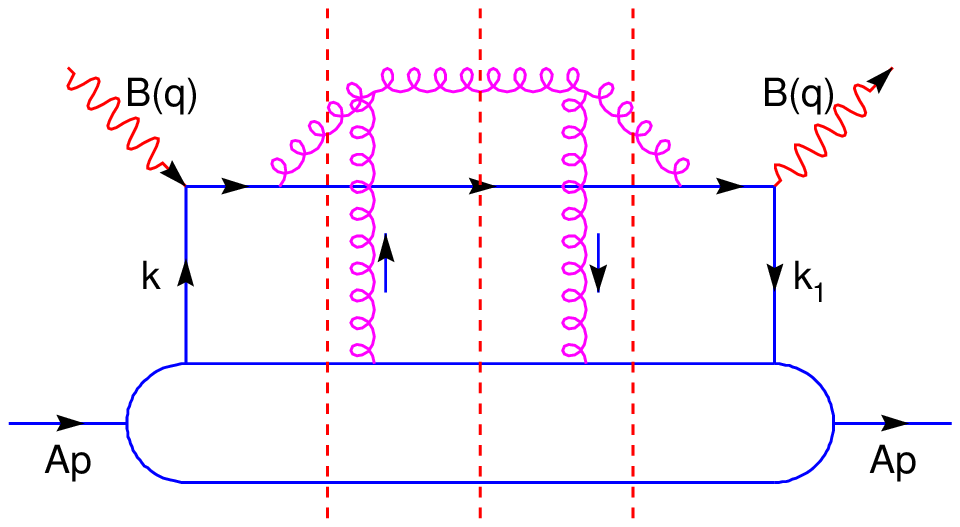,width=2.6in,height=1.8in}}
\caption{ } \label{fig7}
\end{figure}

\begin{figure}
\centerline{\psfig{file=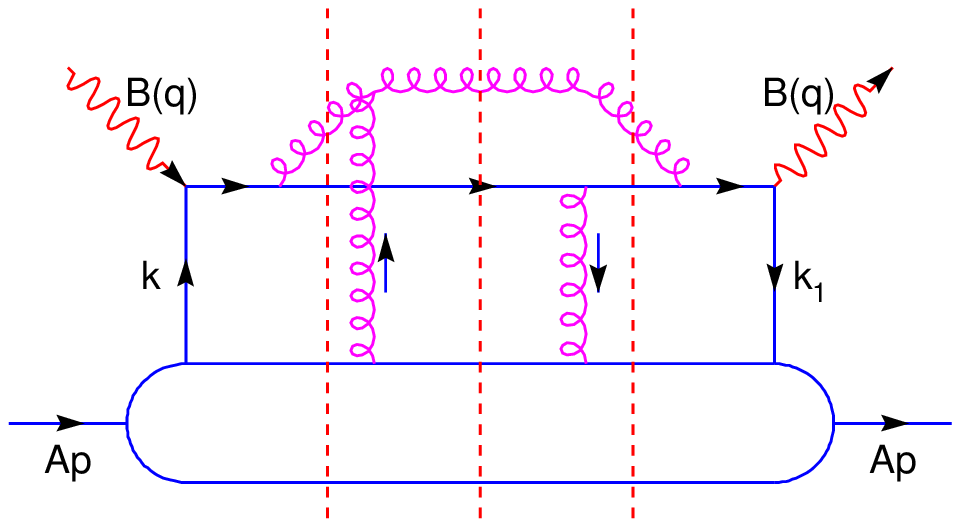,width=2.6in,height=1.8in}}
\caption{ } \label{fig8}
\end{figure}

\begin{figure}
\centerline{\psfig{file=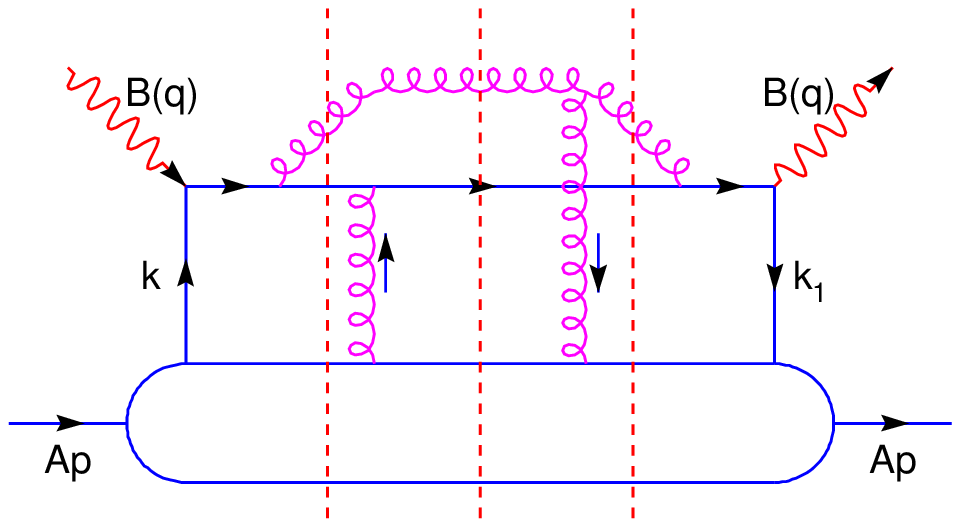,width=2.6in,height=1.8in}}
\caption{ } \label{fig9}
\end{figure}

\begin{figure}
\centerline{\psfig{file=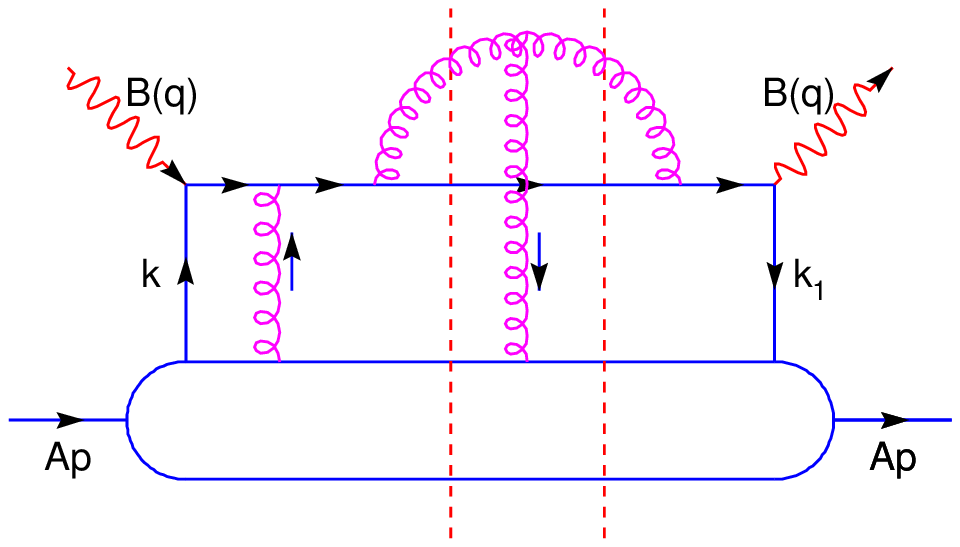,width=2.6in,height=1.8in}}
\caption{   } \label{fig10}
\end{figure}

\begin{figure}
\centerline{\psfig{file=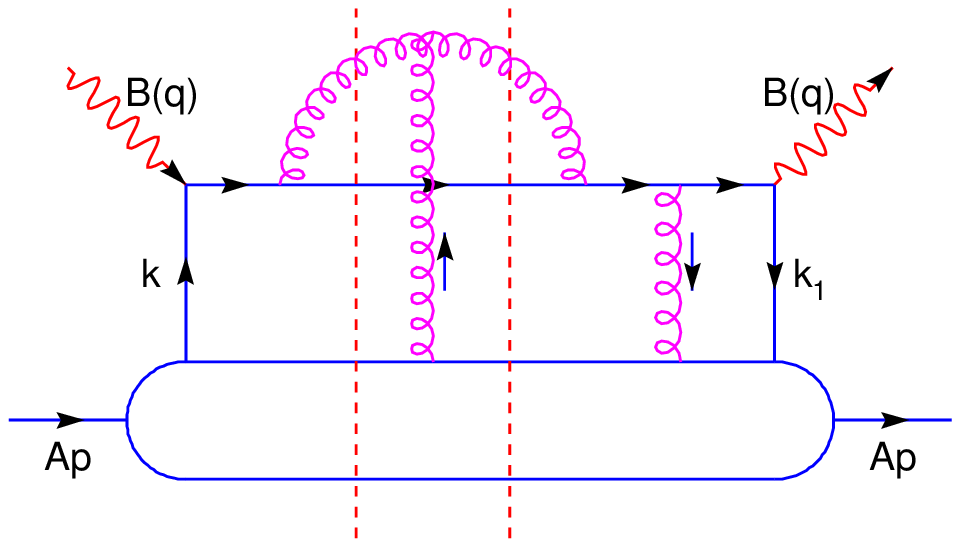,width=2.6in,height=1.8in}}
\caption{   } \label{fig11}
\end{figure}

In this appendix we list the calculation results of double
scattering of the heavy quark discussed in Section III. There are
total 23 cut diagrams as illustrated in
Fig.~\ref{fig01}-\ref{fig11}.

There are three three different cuts(central, left, right) in
Fig.~\ref{fig01}, and their contributions are
\begin{eqnarray}
\overline{H}^D_{1 }(y^-,y_1^-,y_2^-,k_T,x,p,q,M,z)&=& \int
d\ell_T^2\,
\frac{(1+z^2)\ell_T^2+(1-z)^4 M^2}{(1-z)(\ell_T^2+(1-z)^2M^2)^2} \nonumber \\
&\times&\frac{\alpha_s}{2\pi}\, C_F \,
 \frac{2\pi\alpha_s}{N_c}
\overline{I}_{1,C}(y^-,y_1^-,y_2^-,\ell_T,k_T,x,p,q,M,z)
 \, , \label{eq:hc1}
\\
\overline{I}_{1,C }(y^-,y_1^-,y_2^-,\ell_T,k_T,x,p,q,M,z)
&=&e^{i(x+x_L)p^+y^- + ix_Dp^+(y_1^- - y_2^-)}
\theta(-y_2^-)\theta(y^- - y_1^-) \nonumber \\
& &\hspace{-1.0in}\times \left[1-e^{-i(x_L+(1-z)x_M/z
)p^+y_2^-}\right] \left[1-e^{-i(x_L+(1-z)x_M/z)p^+(y^- -
y_1^-)}\right] \; , \label{eq:Ic1}
\\
\overline{I}_{1,L}(y^-,y_1^-,y_2^-,\ell_T,k_T,x,p,q,M,z)\,
&=&-e^{i(x+x_L)p^+y^- + ix_Dp^+(y_1^- - y_2^-)}
\theta(y_1^- - y_2^-)\theta(y^- - y_1^-) \nonumber \\
&\times &(1-e^{-i(x_L+(1-z)x_M/z)p^+(y^- - y_1^-)}) \, ,\label{eq:I1L} \\
\overline{I}_{1,R}(y^-,y_1^-,y_2^-,\ell_T,k_T,x,p,q,M,z)
&=&-e^{i(x+x_L)p^+y^- + ix_Dp^+(y_1^- - y_2^-)}
\theta(-y_2^-)\theta(y_2^- - y_1^-) \nonumber \\
&\times &(1-e^{-i(x_L+(1-z)x_M/z)p^+y_2^-}) \, .\label{eq:I1R}
\end{eqnarray}

In Fig.~\ref{fig2}, there are two different cuts for induced gluon
radiation, central or left. They give,

\begin{eqnarray}
& &\overline{H}^D_{2}(y^-,y_1^-,y_2^-,k_T,x,p,q,M,z) \nonumber \\
& &\hspace{0.5in} =\int d\ell_T^2
\frac{(1+z^2)\vec{\ell_T}\cdot(\vec{\ell_T}-(1-z)\vec{k_T})
+(1-z)^4 M^2} {(1-z)[\ell_T^2 + (1-z)^2 M^2][(
\vec{\ell_T}-(1-z)\vec{k_T})^2 +(1-z)^2 M^2]} \,
 \nonumber \\
& & \hspace{0.5in}\times \frac{\alpha_s}{2\pi}\,
(C_F-\frac{C_A}{2}) \frac{2\pi\alpha_s}{N_c}
\overline{I}_{2}(y^-,y_1^-,y_2^-,\ell_T,k_T,x,p,q,M,z)
 \, , \label{eq:hc2}  \\
\nonumber \\
& &\overline{I}_{2,C}(y^-,y_1^-,y_2^-,\ell_T,k_T,x,p,q,M,z)
= e^{i(x+x_L)p^+y^-+ix_Dp^+(y_1^--y_2^-)}
\theta(-y_2^-)\theta(y^- - y_1^-) \nonumber \\
& & \hspace{0.5in} \times \left[e^{-i(x_L+(1-z)x_M/z)p^+(y^- -
y_1^-)}-e^{-i(x_L+(1-z)x_M/z)p^+(y^- - y_1^-+y_2^-)}\right] \, ,
\label{eq:I2C}  \\
\nonumber \\
& &\overline{I}_{2,L}(y^-,y_1^-,y_2^-,\ell_T,k_T,x,p,q,M,z) e^{i(x+x_L)p^+y^-+ix_Dp^+(y_1^--y_2^-)}
\theta(y^- - y_1^-)\theta(y_1^- - y_2^-) \nonumber \\
& & \hspace{0.5in} \times \left[e^{-i(x_L+(1-z)x_M/z)p^+(y^- -
y_2^-)+i(x_D^0-x_D)p^+(y_1^- - y_2^-)}
 -e^{-i(x_L+(1-z)x_M/z)p^+(y^- - y_1^-)}\right] \, ,
\label{eq:I2L}
\end{eqnarray}
where $x_D^0=k_T^2/2p^+q^-$.

As for the central cut and right cut of Fig.~\ref{fig3}, we obtain
\begin{eqnarray}
 & & \overline{H}^D_{3}(y^-,y_1^-,y_2^-,k_T,x,p,q,M,z) \nonumber \\
& & \hspace{0.5in} =\int d\ell_T^2
\frac{(1+z^2)\vec{\ell_T}\cdot(\vec{\ell_T}-(1-z)\vec{k_T})
+(1-z)^4 M^2} {(1-z)[\ell_T^2 + (1-z)^2 M^2][(
\vec{\ell_T}-(1-z)\vec{k_T})^2 +(1-z)^2 M^2]} \,
 \nonumber \\
& & \hspace{0.5in} \times \frac{\alpha_s}{2\pi}\,
(C_F-\frac{C_A}{2}) \frac{2\pi\alpha_s}{N_c}
\overline{I}_{3}(y^-,y_1^-,y_2^-,\ell_T,k_T,x,p,q,M,z)
 \, , \label{eq:hc3}  \\
\nonumber \\
& &\overline{I}_{3,C}(y^-,y_1^-,y_2^-,\ell_T,k_T,x,p,q,M,z) e^{i(x+x_L)p^+y^-+ix_Dp^+(y_1^--y_2^-)}
\theta(-y_2^-)\theta(y^- - y_1^-) \nonumber \\
& & \hspace{0.5in}\times \left [e^{-i(x_L+(1-z)x_M/z)p^+
y_2^-}-e^{-i(x_L+(1-z)x_M/z)p^+(y^- - y_1^-+y_2^-)}\right] \, ,
\label{eq:I3C}  \\
 \nonumber \\
& &\overline{I}_{3,R}(y^-,y_1^-,y_2^-,\ell_T,k_T,x,p,q,M,z) e^{i(x+x_L)p^+y^-+ix_Dp^+(y_1^--y_2^-)}
\theta(y^- - y_1^-)\theta(y_1^- - y_2^-) \nonumber \\
& & \hspace{0.5in}\times \left
[e^{-i(x_L+(1-z)x_M/z)p^+y_1^-+i(x_D^0-x_D)p^+(y_1^- - y_2^-)}
 -e^{-i(x_L+(1-z)x_M/z)p^+ y_2^-}\right] \, ,
\label{eq:I3R}
\end{eqnarray}

There is only one cut (left cut) in Fig.~\ref{fig4} with the
contribution,
\begin{eqnarray}
\overline{H}^D_{4,C }(y^-,y_1^-,y_2^-,k_T,x,p,q,M,z)&=& \int
d\ell_T^2\,
\frac{(1+z^2)\ell_T^2+(1-z)^4 M^2}{(1-z)[\ell_T^2+(1-z)^2M^2]^2} \nonumber \\
&\times&\frac{\alpha_s}{2\pi}\, C_F \,
 \frac{2\pi\alpha_s}{N_c}
\overline{I}_{4}(y^-,y_1^-,y_2^-,\ell_T,k_T,x,p,q,M,z)
 \, , \label{eq:hc4}   \\
\overline{I}_{4,L}(y^-,y_1^-,y_2^-,\ell_T,k_T,x,p,q,M,z)&=&
-e^{i(x+x_L)p^+y^-+ix_Dp^+(y_1^--y_2^-)}
\theta(y^- - y_1^-)\theta(y_1^- - y_2^-) \nonumber \\
&\times&e^{i(x_D^0-x_D)p^+(y_1^- - y_2^-)}
e^{-i(x_L+(1-z)x_M/z)p^+(y^- - y_2^-)} \, . \label{eq:I4L}
\end{eqnarray}

As for Fig.~\ref{fig5}, we get
\begin{eqnarray}
\overline{H}^D_{5}(y^-,y_1^-,y_2^-,k_T,x,p,q,M,z)&=& \int
d\ell_T^2\, \frac{(1+z^2)(\vec{\ell_T}-(1-z)\vec{k_T})^2+(1-z)^4
M^2}
{(1-z)[(\vec{\ell_T}-(1-z)\vec{k_T})^2 +(1-z)^2 M^2]^2}  \nonumber \\
&\times& \frac{\alpha_s}{2\pi}\, C_F\, \frac{2\pi\alpha_s}{N_c}
\overline{I}_{5}(y^-,y_1^-,y_2^-,\ell_T,k_T,x,p,q,M,z)
 \, , \label{eq:hc5} \\
\overline{I}_{5,C}(y^-,y_1^-,y_2^-,\ell_T,k_T,x,p,q,M,z)&=&
e^{i(x+x_L)p^+y^-+ix_Dp^+(y_1^--y_2^-)}
\theta(-y_2^-)\theta(y^- - y_1^-) \nonumber \\
&\times& e^{-i(x_L+(1-z)x_M/z)p^+(y^- - y_1^- +y_2^-)} \, .
\label{eq:I5C}
\end{eqnarray}

The contribution from Fig.~\ref{fig6} is
\begin{eqnarray}
\overline{H}^D_{6,C }(y^-,y_1^-,y_2^-,k_T,x,p,q,M,z)&=& \int
d\ell_T^2\,
\frac{(1+z^2)\ell_T^2+(1-z)^4 M^2}{(1-z)[\ell_T^2+(1-z)^2M^2]^2} \nonumber \\
&\times&\frac{\alpha_s}{2\pi}\, C_F \,
 \frac{2\pi\alpha_s}{N_c}
\overline{I}_{6}(y^-,y_1^-,y_2^-,\ell_T,k_T,x,p,q,M,z)
 \, , \label{eq:hc6}   \\
\overline{I}_{6,R}(y^-,y_1^-,y_2^-,\ell_T,k_T,x,p,q,M,z)&=&
-e^{i(x+x_L)p^+y^-+ix_Dp^+(y_1^--y_2^-)}
\theta(-y_2^-)\theta(y_2^- - y_1^-) \nonumber \\
&\times&e^{i(x_D^0-x_D)p^+(y_1^- -
y_2^-)}e^{-i(x_L+(1-z)x_M/z)p^+y_1^-}
 \, .
\label{eq:I6R}
\end{eqnarray}


As for the processes in Fig.~\ref{fig7} we have three possible
cuts with the corresponding contributions,
\begin{eqnarray}
\overline{H}^D_{7,C}(y^-,y_1^-,y_2^-,k_T,x,p,q,M,z)  &=&\int
d\ell_T^2 \frac{(1+z^2)(\vec{\ell_T}-\vec{k_T})^2 +(1-z)^4 M^2}
{(1-z)[( \vec{\ell_T}-\vec{k_T})^2 +(1-z)^2 M^2]^2} \,
 \nonumber \\ &\times&\frac{\alpha_s}{2\pi}\,
C_A\, \frac{2\pi\alpha_s}{N_c}
\overline{I}_{7,C}(y^-,y_1^-,y_2^-,\ell_T,k_T,x,p,q,M,z)
 \, , \label{eq:hc7-1}  \\
\overline{I}_{7,C}(y^-,y_1^-,y_2^-,\ell_T,k_T,x,p,q,M,z) &=&
e^{i(x+x_L)p^+y^-+ix_Dp^+(y_1^--y_2^-)}
\theta(-y_2^-)\theta(y^- - y_1^-) \nonumber \\
&\times&\left [e^{ix_Dp^+y_2^-/(1-z)}
  -e^{-i(x_L+(1-z)x_M/z)p^+y_2^-}\right] \nonumber \\
&\times&\left[e^{ix_Dp^+(y^- - y_1^-)/(1-z)}
-e^{-i(x_L+(1-z)x_M/z)p^+(y^- - y_1^-)}\right] \, , \label{eq:I7C} \\
\nonumber \\
\overline{H}^D_{7,L(R)}(y^-,y_1^-,y_2^-,k_T,x,p,q,M,z) &=&\int
d\ell_T^2 \frac{(1+z^2)\ell_T^2 +(1-z)^4 M^2} {(1-z)[\ell_T^2 +
(1-z)^2 M^2]^2} \,
 \nonumber \\ &\times&\frac{\alpha_s}{2\pi}\,
C_A\, \frac{2\pi\alpha_s}{N_c}
\overline{I}_{7,L(R)}(y^-,y_1^-,y_2^-,\ell_T,k_T,x,p,q,M,z)
 \, , \label{eq:hc7-2}  \\
\overline{I}_{7,L}(y^-,y_1^-,y_2^-,\ell_T,k_T,x,p,q,M,z)&=&
-e^{i(x+x_L)p^+y^-+ix_Dp^+(y_1^--y_2^-)}
\theta(y^- - y_1^-)\theta(y_1^- - y_2^-) \nonumber \\
&\times& e^{-i(1-z/(1-z) )x_Dp^+(y_1^- - y_2^-)} \nonumber \\
&\times&\left[1-e^{-i(x_L+(1-z)x_M/z)p^+(y^- - y_1^-)}\right] \, ,
\label{eq:I7L}  \\
\overline{I}_{7,R}(y^-,y_1^-,y_2^-,\ell_T,k_T,x,p,q,M,z)&=&
-e^{i(x+x_L)p^+y^-+ix_Dp^+(y_1^--y_2^-)}
\theta(-y_2^-)\theta(y_2^- - y_1^-) \nonumber \\
&\times& e^{-i(1-z/(1-z) )x_Dp^+(y_1^- - y_2^-)}
\left[1-e^{-i(x_L+(1-z)x_M/z)p^+y_2^-}\right] \, . \label{eq:I7R}
\end{eqnarray}

There are also three cuts in Fig.~\ref{fig8} and we have

\begin{eqnarray}
& &\overline{H}^D_{8}(y^-,y_1^-,y_2^-,k_T,x,p,q,M,z) =\int
d\ell_T^2 \frac{(1+z^2)\vec{\ell_T}\cdot(\vec{\ell_T}-\vec{k_T})
+(1-z)^4 M^2} {(1-z)[\ell_T^2 + (1-z)^2 M^2][(
\vec{\ell_T}-\vec{k_T})^2 +(1-z)^2 M^2]} \,
 \nonumber \\
& & \hspace{0.5in} \times \frac{\alpha_s}{2\pi}\, \frac{C_A}{2}\,
\frac{2\pi\alpha_s}{N_c}
\overline{I}_{8}(y^-,y_1^-,y_2^-,\ell_T,k_T,x,p,q,M,z)
 \, , \label{eq:hc8}
\end{eqnarray}
\begin{eqnarray}
\overline{I}_{8,C}(y^-,y_1^-,y_2^-,\ell_T,k_T,x,p,q,M,z)&=&
-e^{i(x+x_L)p^+y^-+ix_Dp^+(y_1^--y_2^-)}
\theta(-y_2^-)\theta(y^- - y_1^-) \nonumber \\
&\times&
\left[e^{ix_Dp^+y_2^-/(1-z)}-e^{-i(x_L+(1-z)x_M/z)p^+y_2^-}\right] \nonumber \\
&\times&\left[1-e^{-i(x_L+(1-z)x_M/z)p^+(y^- - y_1^-)}\right] \, ,
\label{eq:I8C}  \\
\overline{I}_{8,L}(y^-,y_1^-,y_2^-,\ell_T,k_T,x,p,q,M,z)&=&
-e^{i(x+x_L)p^+y^-+ix_Dp^+(y_1^--y_2^-)}
\theta(y^- - y_1^-)\theta(y_1^- - y_2^-) \nonumber \\
&\times& e^{-i(1-z/(1-z))x_Dp^+(y_1^- - y_2^-)} \nonumber \\
&\times&\left[e^{-i(x_L+(1-z)x_M/z)p^+(y^- - y_1^-)}
-e^{ix_Dp^+(y^- - y_1^-)/(1-z)}\right] \, ,
\label{eq:I8L}  \\
\overline{I}_{8,R}(y^-,y_1^-,y_2^-,\ell_T,k_T,x,p,q,M,z)&=&
-e^{i(x+x_L)p^+y^-+ix_Dp^+(y_1^--y_2^-)}
\theta(-y_2^-)\theta(y_2^- - y_1^-) \nonumber \\
&\times&\left[e^{-i(x_L+(1-z)x_M/z)p^+y_2^-}
-e^{ix_Dp^+y_2^-/(1-z)}\right] \, . \label{eq:I8R}
\end{eqnarray}

The contributions from the three cuts in Fig.~\ref{fig9} are
\begin{eqnarray}
& &\overline{H}^D_{9}(y^-,y_1^-,y_2^-,k_T,x,p,q,M,z) =\int
d\ell_T^2 \frac{(1+z^2)\vec{\ell_T}\cdot(\vec{\ell_T}-\vec{k_T})
+(1-z)^4 M^2} {(1-z)[\ell_T^2 + (1-z)^2 M^2][(
\vec{\ell_T}-\vec{k_T})^2 +(1-z)^2 M^2]} \,
 \nonumber \\
& & \hspace{0.5in} \times \frac{\alpha_s}{2\pi}\, \frac{C_A}{2}\,
\frac{2\pi\alpha_s}{N_c}
\overline{I}_{9}(y^-,y_1^-,y_2^-,\ell_T,k_T,x,p,q,M,z)
 \, , \label{eq:hc9}
\end{eqnarray}

\begin{eqnarray}
\overline{I}_{9,C}(y^-,y_1^-,y_2^-,\ell_T,k_T,x,p,q,M,z)&=&
-e^{i(x+x_L)p^+y^-+ix_Dp^+(y_1^--y_2^-)}
\theta(-y_2^-)\theta(y^- - y_1^-) \nonumber \\
&\times&\left[e^{ix_Dp^+(y^- -
y_1^-)/(1-z)}-e^{-i(x_L+(1-z)x_M/z)p^+(y^- - y_1^-)}\right]
\nonumber \\
&\times&\left[1-e^{-i(x_L+(1-z)x_M/z)p^+ y_2^-}\right] \, ,
\label{eq:I9C}  \\
\overline{I}_{9,L}(y^-,y_1^-,y_2^-,\ell_T,k_T,x,p,q,M,z)&=&
-e^{i(x+x_L)p^+y^-+ix_Dp^+(y_1^--y_2^-)}
\theta(y^- - y_1^-)\theta(y_1^- - y_2^-) \nonumber \\
&\times&\left[e^{-i(x_L+(1-z)x_M/z)p^+ (y^- - y_1^-)}
-e^{ix_Dp^+(y^- - y_1^-)/(1-z)}\right] \, ,
\label{eq:I9L}   \\
\overline{I}_{9,R}(y^-,y_1^-,y_2^-,\ell_T,k_T,x,p,q,M,z)&=&
-e^{i(x+x_L)p^+y^-+ix_Dp^+(y_1^--y_2^-)}
\theta(-y_2^-)\theta(y_2^- - y_1^-) \nonumber \\
&\times& e^{-i(1-z/(1-z))x_Dp^+(y_1^- - y_2^-)} \nonumber \\
&\times&\left[e^{-i(x_L+(1-z)x_M/z)p^+y_2^-}
-e^{ix_Dp^+y_2^-/(1-z)}\right] \, , \label{eq:I9R}
\end{eqnarray}

>From Fig.~\ref{fig10} with two possible cuts (central or left), we
have
\begin{eqnarray}
& &\overline{H}^D_{10,C}(y^-,y_1^-,y_2^-,k_T,x,p,q,M,z) \nonumber \\
& & \hspace{0.5in} =\int d\ell_T^2
\frac{(1+z^2)(\vec{\ell_T}-\vec{k_T})\cdot(\vec{\ell_T}-(1-z)\vec{k_T})
+(1-z)^4 M^2} {(1-z)[(\vec{\ell_T}-\vec{k_T})^2 + (1-z)^2 M^2][(
\vec{\ell_T}-(1-z)\vec{k_T})^2 +(1-z)^2 M^2]} \,
 \nonumber \\
& &\hspace{0.5in} \times \frac{\alpha_s}{2\pi}\, \frac{C_A}{2}\,
\frac{2\pi\alpha_s}{N_c}
\overline{I}_{10,C}(y^-,y_1^-,y_2^-,\ell_T,k_T,x,p,q,M,z)
 \, , \label{eq:hc10-1}  \\
& & \overline{I}_{10,C}(y^-,y_1^-,y_2^-,\ell_T,k_T,x,p,q,M,z) e^{i(x+x_L)p^+y^-+ix_Dp^+(y_1^--y_2^-)} \,
\theta(-y_2^-)\theta(y^- - y_1^-) \nonumber \\
& & \hspace{0.5in} \times e^{-i(x_L+(1-z)x_M/z)p^+y_2^-}
\left[e^{ix_Dp^+(y^- - y_1^-)/(1-z)}-e^{-i(x_L+(1-z)x_M/z)p^+(y^-
- y_1^-)}\right] \, ,
\label{eq:I10C}  \\ \nonumber \\
& &\overline{H}^D_{10,R}(y^-,y_1^-,y_2^-,k_T,x,p,q,M,z) \nonumber \\
& & \hspace{0.5in}= \int d\ell_T^2
\frac{(1+z^2)\vec{\ell_T}\cdot(\vec{\ell_T}-z\vec{k_T}) +(1-z)^4
M^2} {(1-z)[\ell_T^2 + (1-z)^2 M^2][( \vec{\ell_T}-z\vec{k_T})^2
+(1-z)^2 M^2]} \,
 \nonumber \\
& & \hspace{0.5in} \times  \frac{\alpha_s}{2\pi}\, \frac{C_A}{2}\,
\frac{2\pi\alpha_s}{N_c}
\overline{I}_{10,R}(y^-,y_1^-,y_2^-,\ell_T,k_T,x,p,q,M,z)
 \, , \label{eq:hc10-2}  \\
& &\overline{I}_{10,R}(y^-,y_1^-,y_2^-,\ell_T,k_T,x,p,q,M,z)
=e^{i(x+x_L)p^+y^-+ix_Dp^+(y_1^--y_2^-)} \,
\theta(-y_2^-)\theta(y_2^- - y_1^-) \nonumber \\
& & \hspace{0.5in} \times \left[e^{-i(x_D-x_D^0)p^+(y_1^- -
y_2^-)-i(x_L+(1-z)x_M/z)p^+y_1^-} \right.
\nonumber  \\
& & \hspace{1.0in}\left. -e^{-i(1-z/(1-z))x_D p^+(y_1^- - y_2^-)
-i(x_L+(1-z)x_M/z)p^+y_2^-}\right] \, . \label{eq:I10R}
\end{eqnarray}

Finally, Fig.~\ref{fig11} has both central cut and the left cut
with contributions as
\begin{eqnarray}
& &\overline{H}^D_{11,C}(y^-,y_1^-,y_2^-,k_T,x,p,q,M,z) \nonumber \\
& & \hspace{0.5in} =\int d\ell_T^2
\frac{(1+z^2)(\vec{\ell_T}-\vec{k_T})\cdot(\vec{\ell_T}-(1-z)\vec{k_T})
+(1-z)^4 M^2} {(1-z)[(\vec{\ell_T}-\vec{k_T})^2 + (1-z)^2 M^2][(
\vec{\ell_T}-(1-z)\vec{k_T})^2 +(1-z)^2 M^2]} \,
 \nonumber \\
&  & \hspace{0.5in} \times \frac{\alpha_s}{2\pi}\, \frac{C_A}{2}\,
\frac{2\pi\alpha_s}{N_c}
\overline{I}_{11,C}(y^-,y_1^-,y_2^-,\ell_T,k_T,x,p,q,M,z)
 \, , \label{eq:hc11-1}  \\
& & \overline{I}_{11,C}(y^-,y_1^-,y_2^-,\ell_T,k_T,x,p,q,M,z) e^{i(x+x_L)p^+y^-+ix_Dp^+(y_1^--y_2^-)} \,
\theta(-y_2^-)\theta(y^- - y_1^-) \nonumber \\
&& \hspace{0.5in}\times e^{-i(x_L+(1-z)x_M/z)p^+(y^- - y_1^-)}
\left[e^{ix_Dp^+ y_2^-/(1-z)}-e^{-i(x_L+(1-z)x_M/z)p^+
y_2^-}\right] \, ,
\label{eq:I11C}  \\  \nonumber \\
& &\overline{H}^D_{11,L}(y^-,y_1^-,y_2^-,k_T,x,p,q,M,z) \nonumber \\
&& \hspace{0.5in}=\int d\ell_T^2
\frac{(1+z^2)\vec{\ell_T}\cdot(\vec{\ell_T}-z\vec{k_T}) +(1-z)^4
M^2} {(1-z)[\ell_T^2 + (1-z)^2 M^2][( \vec{\ell_T}-z\vec{k_T})^2
+(1-z)^2 M^2]} \,
 \nonumber \\
&& \hspace{0.5in}\times  \frac{\alpha_s}{2\pi}\, \frac{C_A}{2}\,
\frac{2\pi\alpha_s}{N_c}
\overline{I}_{11,L}(y^-,y_1^-,y_2^-,\ell_T,k_T,x,p,q,M,z)
 \, , \label{eq:hc11-2}  \\
& &\overline{I}_{11,L}(y^-,y_1^-,y_2^-,\ell_T,k_T,x,p,q,M,z) e^{i(x+x_L)p^+y^-+ix_Dp^+(y_1^--y_2^-)} \,
\theta(y^- - y_1^-)\theta(y_1^- - y_2^-) \nonumber \\
& & \hspace{0.5in} \times \left[e^{-i(x_D-x_D^0)p^+(y_1^- -
y_2^-)-i(x_L+(1-z)x_M/z)p^+(y^- - y_2^-)}\right.
\nonumber  \\
& & \hspace{0.5in}\left.-e^{-i(1-z/(1-z))x_D p^+(y_1^- - y_2^-)
-i(x_L+(1-z)x_M/z)p^+(y^- - y_1^-)}\right] \, . \label{eq:I11L}
\end{eqnarray}


\end{document}